\documentclass[sigconf]{acmart}
\renewcommand\footnotetextcopyrightpermission[1]{} % removes footnote with conference information in first column
\usepackage{tugraz_defaults}
\usepackage{balance} 
\graphicspath{{figures/}}
\DeclareGraphicsExtensions{.pdf,.jpeg,.png}

\begin{document}
\fancyhead{}

\newcommand{\riscv}{\mbox{RISC-V}\xspace}
\newcommand{\tagenc}{CrypTag\xspace}
\newcommand{\mstp}{\texttt{mstp}\xspace}
\newcommand{\tilelink}{TileLink\xspace}
\newcommand{\memsec}{\mbox{\textsc{MEMSEC}}\xspace}
\newcommand{\tauth}{\textbf{S1}\xspace}
\newcommand{\tenc}{\textbf{S2}\xspace}

\def\ci#1{\textcircled{\resizebox{.5em}{!}{#1}}}
\makeatletter
\renewcommand\paragraph{\@startsection{paragraph}{4}{\z@}%
                                      {\parskip}%{3.25ex \@plus1ex \@minus.2ex}%
                                      {-0.5em}%
                                      {\normalfont\normalsize\bfseries}}

\title{\tagenc: Thwarting Physical and Logical Memory Vulnerabilities using Cryptographically Colored Memory}

\author{Pascal Nasahl}
\email{pascal.nasahl@iaik.tugraz.at}
\affiliation{%
  \institution{Graz University of Technology}}

\author{Robert Schilling}
\email{robert.schilling@iaik.tugraz.at}
\affiliation{%
  \institution{Graz University of Technology}}

\author{Mario Werner}
\email{mario.werner@iaik.tugraz.at}
\affiliation{%
  \institution{Graz University of Technology}}

\author{Jan Hoogerbrugge}
\email{jan.hoogerbrugge@nxp.com}
\affiliation{%
  \institution{NXP Semiconductors Netherlands}}

\author{Marcel Medwed}
\email{marcel.medwed@nxp.com}
\affiliation{%
  \institution{NXP Semiconductors Austria}}

\author{Stefan Mangard}
\email{stefan.mangard@iaik.tugraz.at}
\affiliation{%
  \institution{Graz University of Technology}
  \institution{Lamarr Security Research}}

\renewcommand{\shortauthors}{Nasahl, et al.}

\begin{abstract}

Memory vulnerabilities are a major threat to many computing systems.
To effectively thwart spatial and temporal memory vulnerabilities, full logical memory safety is required.
However, current mitigation techniques for memory safety are either too expensive or trade security against efficiency.
One promising attempt to detect memory safety vulnerabilities in hardware is memory coloring, a security policy deployed on top of tagged memory architectures.
However, due to the memory storage and bandwidth overhead of large tags, commodity tagged memory architectures usually only provide small tag sizes, thus limiting their use for security applications.

Irrespective of logical memory safety, physical memory safety is a necessity in hostile environments prevalent for modern cloud computing and IoT devices.
Architectures from Intel and AMD already implement transparent memory encryption to maintain confidentiality and integrity of all off-chip data.
Surprisingly, the combination of both, logical and physical memory safety, has not yet been extensively studied in previous research, and a na\"ive combination of both security strategies would accumulate both overheads.

In this paper, we propose \tagenc, an efficient hardware/software co-design mitigating a large class of logical memory safety issues and providing full physical memory safety.
At its core, \tagenc utilizes a transparent memory encryption engine not only for physical memory safety, but also for memory coloring at hardly any additional costs.
The design avoids any overhead for tag storage by embedding memory colors in the upper bits of a pointer and using these bits as an additional input for the memory encryption.
A custom compiler extension automatically leverages \tagenc to detect logical memory safety issues for commodity programs and is fully backward compatible.

For evaluating the design, we extended a \riscv processor with memory encryption with \tagenc.
Furthermore, we developed a LLVM-based toolchain automatically protecting all dynamic, local, and global data.
Our evaluation shows a hardware overhead of less than 1\,\% and an average runtime overhead between 1.5\,\% and 6.1\,\% for thwarting logical memory safety vulnerabilities on a system already featuring memory encryption.
Enhancing a system with memory encryption typically induces a runtime overhead between 5\,\% and 109.8\,\% for commercial and open-source encryption units.

\end{abstract}


\maketitle

\section{Introduction}
\label{sec:intro}
According to MITRE~\cite{Mitre2019}, three out of ten of the most common software weaknesses in 2019 leading to security vulnerabilities are owed to logical memory safety violations.
Memory vulnerabilities, which exploit spatial or temporal memory bugs, are the foundation for more sophisticated attack techniques, such as return-oriented programming~(ROP)~\cite{DBLP:conf/ccs/Shacham07} or data-oriented programming~(DOP)~\cite{DBLP:conf/sp/HuSACSL16}.
Consequently, due to the high impact of memory vulnerabilities, defense strategies, such as code- and data-pointer integrity~\cite{DBLP:books/mc/18/KuznetsovSPCSS18,DBLP:conf/uss/LiljestrandNWPE19} or the protection of the control-flow~\cite{DBLP:conf/ccs/MashtizadehBBM15}, were introduced in the past.
However, these concepts only complicate the exploitation of a memory vulnerability, but do not fix the root cause.
To completely thwart memory vulnerabilities, full logical memory safety for all classes of memory allocations is necessary~\cite{DBLP:conf/sp/SzekeresPWS13}.
Unfortunately, software solutions providing memory safety, such as SoftBound~\cite{DBLP:conf/pldi/NagarakatteZMZ09} or CETS~\cite{DBLP:conf/iwmm/NagarakatteZMZ10}, typically add significant overhead and increase costs if deployed on a larger scale.
High performance penalty can be counteracted with hardware assistance.
A promising attempt to detect memory safety violations with hardware support are tagged memory architectures~\cite{DBLP:journals/usenix-login/Serebryany19}.
Tagged memory assigns additional metadata to the memory, enforcing different security policies~\cite{DBLP:conf/osdi/ZeldovichKDK08, DBLP:conf/isca/WoodruffWCMADLNNR14}.
One policy, allowing to detect memory safety vulnerabilities, is memory coloring, which is implemented on top of tagged memory.
The basic idea of memory coloring is to lock each memory allocation through a key.
A later memory access is only permitted when using the correct key.
This \textit{lock-and-key} approach is implemented in the ARM Memory Tagging Extension~(MTE)~\cite{ARM2019} and provides tagged memory in hardware.
Google announced to work on deploying memory coloring based on MTE in Android on a larger scale~\cite{Serebryany2019}, through the MemTagSanitizer~\cite{MemTagSan2020} project integrated into LLVM~\cite{DBLP:conf/cgo/LattnerA04}.
While this concept is a step in the right direction, the memory overhead for storing the tags is still problematic for large-scale applications.
To reduce the memory overhead of memory coloring, ARM decided to limit their concept to small tags.
In ARM MTE, a 16 byte memory block is tagged with a \SI{4}{\bits} tag, resulting in a memory overhead of 3.125\,\%.
While this memory overhead might be feasible for most applications, a tag size of \SI{4}{\bits} only leads to 16 individual colors, thus limiting the use of MTE as a security mechanism and making debugging the main application possible.
Increasing the tag size from 4 to \SI{16}{\bits} not only increases the available color space and, therefore, also the security guarantees, but also raises the memory overhead to 12.5\,\%.
\enlargethispage*{5mm}
In addition to logical memory violations, systems deployed in hostile environments, such as cloud services or IoT devices, need to consider physical attacks on the system memory in their threat model.
To maintain confidentiality and integrity of data stored in off-chip memory, memory encryption is a widely used technique.
Although deploying transparent memory encryption is costly, vendors like Intel and AMD acknowledge the significant threat of physical attacks and offer schemes like Software Guard Extensions technology~(SGX)~\cite{2016sgx}, Multi-Key Total Memory Encryption~(MKTME)~\cite{Corporation2019}, or Transparent Secure Memory Encryption~(TSME)~\cite{Kaplan2016a} for the consumer market.
Memory encryption is also employed on smaller devices, \eg in the Internet-of-Things~(IoT), to protect sensitive data in memory~\cite{DBLP:journals/corr/abs-1907-10119,DBLP:journals/jce/UnterluggauerWM19}.

Despite the immense threat of logical and physical memory safety violations, the efficient combination of both mitigation strategies has not yet been extensively studied in past research and a na\"ive combination of both security strategies would accumulate both overheads.

\subsection*{Contribution}
In our paper, we introduce \tagenc, a hardware/software co-design mitigating a broad range of logical memory safety issues and providing full physical memory safety.
We demonstrate that realizing memory coloring on top of an already implemented memory encryption unit hardly costs more.
In exploiting properties of the memory encryption scheme, we overcome limitations of traditional memory coloring schemes.
While previous tagged memory architectures~\cite{ARM2019,DBLP:journals/micro/AingaranJKLLMPR15,DBLP:conf/osdi/ZeldovichKDK08,song2015towards, DBLP:conf/iccd/JoannouWKMBXWCR17, DBLP:conf/sp/SongMAYLKLP16} store the tag in memory and trade security against lower memory overhead, \tagenc completely avoids storing tags in memory and thus allows using larger tag sizes.
\tagenc uses the memory color as additional input for the memory encryption scheme to encrypt every allocation differently.
Inside the processor, we store the color information directly in the upper bits of the pointer, avoiding any additional storage overhead there.
The tag is propagated through the system, stored in the cache, and finally used to encrypt the data when being stored in memory.
Based on the capabilities of the underlying memory encryption engine, we derive two security policies for \tagenc.

We further present a software concept utilizing the hardware architecture to mitigate memory safety violations.
Our approach assigns each allocated memory object on the heap, stack, and global data a random color.
When accessing a memory object with the wrong color, \eg due to a spatial or temporal memory bug, the \tagenc architecture, in its strongest security policy, identifies the color mismatch.
This strategy allows us to successfully detect most spatial and temporal memory vulnerabilities, enhancing logical memory safety.

To evaluate our concept, we implemented an FPGA prototype based on the \riscv CVA6 core.
Furthermore, we extended the LLVM compiler to automatically instrument the code and protect all memory allocations without the need for user annotations.
We evaluate the performance of \tagenc by executing different programs, from microbenchmarks to application code on our FPGA-based prototyping platform with Linux as host operating system.
The evaluation shows that the performance penalty introduced by \tagenc is less than 6.1\,\% on a system already featuring a memory encryption engine.
Summarized, our contributions are:

\paragraph{Memory Coloring Hardware Architecture:}
We efficiently combine memory encryption with memory coloring and show that the overhead for storing tags in memory can be entirely eliminated.
This allows us to scale the tag size without additional memory cost.

\paragraph{Memory Safety Concept:}
We develop a hardware-assisted memory safety concept based on our memory coloring architecture.
We demonstrate that the increased tag size of \tagenc achieves stronger security guarantees than comparable hardware-assisted memory safety designs, such as ARM MTE.

\paragraph{Prototype Implementation:}
We extend the \riscv CVA6 core with a memory encryption engine and our memory coloring approach.
We further provide a modified LLVM-based toolchain enforcing hardware-assisted memory safety by automatically instrumenting the application code.
We show that the hardware overhead, for a system already using a memory encryption scheme, is less than 1\,\% and the software overhead is between 1.5\,\% and 6.1\,\%.
While highly optimized commercial memory encryption systems typically induce an overhead between 5\,\% and 26\,\%~\cite{ARMMENC}, our evaluation of an open-source memory encryption unit shows a runtime overhead between 58.9\,\% and 109.8\,\%.

\section{Background}
\label{sec:background}

In this section, we discuss the backgrounds of memory safety, tagged memory, and memory encryption in general.

\subsection{Memory Safety}
According to Microsoft~\cite{Microsoft19}, $70\,\%$ of all security bugs fixed in Microsoft products are related to memory safety.
Most of these bugs are critical because they could serve as an entry point for various other attacks.
These attack techniques either tamper the control-flow or the data-flow of a program.
To limit the impact of exploitable memory safety bugs, several attack mitigations like W$\oplus$X or DEP are deployed in modern computer architectures.
However, these countermeasures typically only raise the bar for a successful attack.
Although simpler attacks, like the execution of attacker-injected code, can be mitigated, more advanced techniques, such as ROP, still can bypass these protection mechanisms~\cite{DBLP:conf/ccs/Shacham07}.
Even more sophisticated countermeasures, like ensuring the integrity of the control-flow~\cite{DBLP:conf/ccs/MashtizadehBBM15}, can be defeated by techniques like DOP~\cite{DBLP:conf/sp/HuSACSL16}, where an attacker can craft turing-complete exploits.
To successfully defeat memory vulnerabilities, memory safety is required~\cite{DBLP:conf/sp/SzekeresPWS13}.
Memory safety can be achieved by preventing all spatial and temporal memory vulnerabilities in the system.
A spatial error is classified as dereferencing an out-of-bound pointer, such as accessing an array beyond its bounds, \eg on the stack.
A temporal error, in contrast, occurs when dereferencing a pointer to an already deallocated memory object~\cite{xu2004efficient}.
In the past, memory safety concepts have already been presented.
Watchdog~\cite{DBLP:conf/isca/NagarakatteMZ12}, a hardware-based temporal memory protection scheme, assigns metadata to each allocated object and modifies this metadata on each memory deallocation.
In comparing this metadata with the identifier stored in the pointer, potential temporal memory violations can be detected.
Watchdog can also be extended to find spatial memory bugs.
In addition, SoftBound~\cite{DBLP:conf/pldi/NagarakatteZMZ09} assures spatial memory safety by storing the memory bounds of objects in a shadow memory.
Because of the expensive monitoring of the object bounds by software checks, SoftBound adds an average runtime overhead of around $67\,\%$.
By combining SoftBound with CETS~\cite{DBLP:conf/iwmm/NagarakatteZMZ10}, temporal memory safety can be guaranteed, leading to full memory safety.
However, the large performance overhead of $116\,\%$ on average makes the deployment hard on a larger scale.

\subsection{Tagged Memory}
The concept of tagged memory~\cite{DBLP:conf/afips/Feustel72,Mayer1982,DBLP:journals/ibmsj/ClarkC89} is long-established and describes the idea of associating blocks of memory with additional metadata, \ie \textit{tags}.
Particularly, $TG$-bytes of memory are linked with a $TS$-bits wide tag, where $TG$ denotes the tag granularity and $TS$ the tag size.
In these early computer architectures, tags were primarily used for debugging and for dynamically tracking the numeric type of data words.
However, since tag bits are only memory, somewhat arbitrary policies can be implemented~\cite{DBLP:conf/hpca/VenkataramaniDSP08,DBLP:conf/isca/DhawanVRCSKPD14}.
\begin{figure}[t]
  \begin{center}
    \includegraphics[width=0.8\linewidth]{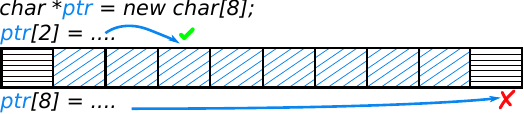}
    \caption{Memory coloring enforcing spatial memory safety.}
    \label{fig:mem_col}
  \end{center}
\end{figure}
Many recent designs utilize tags primarily for memory coloring, as shown in Figure~\ref{fig:mem_col}.
In such a coloring scheme, specific tag values, denoted as colors, are assigned to larger memory regions.
When accessing the memory, these colors are used to determine if a particular read or write operation is genuine.
A mismatch between the color of the accessed memory and the expected color results in a memory error.
Memory coloring is used, \eg for debugging~\cite{DBLP:conf/usenix/SerebryanyBPV12}, isolation~\cite{DBLP:conf/isca/WoodruffWCMADLNNR14}, access control~\cite{DBLP:conf/sp/SongMAYLKLP16,DBLP:conf/ndss/WeiserWBMMS19}, and for enforcing memory safety~\cite{DBLP:journals/corr/abs-1802-09517}.
With ARM's new Armv8.5-A instruction set, the Memory Tagging Extension (MTE)~\cite{ARM2019} was announced, which embeds a tagged memory architecture into consumer hardware, such as mobile phones.
A first attempt using the tagged memory approach on a larger scale is already integrated into the MemTagSanitizer~\cite{MemTagSan2020} project of LLVM.
Similar to the address sanitizer ASan~\cite{ASAN2020} and the hardware-assisted address sanitizer HWASAN~\cite{HWASAN2020}, Google's MemTagSanitizer intends to detect several spatial and temporal memory bugs.
As the MemTagSanitizer benefits from hardware features, the high performance overhead of comparable software-based address sanitizer solutions is reduced to a minimum.
Nevertheless, MTE requires the architecture to store the tags in memory.
To avoid large memory overheads, MTE uses a small tag size of \SI{4}{\bits}, resulting in only 16 distinct memory colors.
However, the security of the memory coloring scheme directly depends on the number of unique colors.
Since colors are assigned randomly for each memory object, two adjacent objects can have the same color.
For security critical systems, a detection probability of only 93.7\,\%, when having a tag size of \SI{4}{\bits}, is insufficient.
Increasing the tag size from 4 to \SI{16}{\bits} would already result in a detection probability of 99.998\,\%, but also increases the memory overhead for tag storage from 3.125\,\% to 12.5\,\%.

While the Armv8.5-A architecture with the MTE feature has not yet been released in hardware, SPARC already implements a hardware-based memory tagging scheme with the Application Data Integrity (ADI)~\cite{DBLP:journals/micro/AingaranJKLLMPR15} feature embedded into Oracle's SPARC M7 processor.
Similar to ARM MTE, the SPARC ADI feature also only supports a tag size of \SI{4}{\bits}.

\subsection{Memory Encryption and Authentication}
\label{sec:memconcept}
Memory safety does not protect the system from physical attacks, such as cold-boot attacks~\cite{DBLP:conf/uss/HaldermanSHCPCFAF08} or RowHammer~\cite{DBLP:conf/isca/KimDKFLLWLM14}.
To counteract these attacks, CPU vendors like Intel and AMD deployed memory encryption into their systems.
Two strategies of transparent memory encryption/authentication are already widely used.
The first and most common variant solely performs encryption to achieve confidentiality.
Examples for memory encryption schemes are Intel's Total Memory Encryption (TME)~\cite{Corporation2019} and AMD's Secure Memory Encryption (SME)~\cite{Kaplan2016a}.
The advantage of these schemes is the high performance and the lack of memory overhead.
The second variant is based on Authenticated Encryption~(AE) and provides both, confidentiality and authenticity, as does the encryption in Intel's Software Guard Extensions~(SGX)~\cite{DBLP:journals/iacr/Gueron16}.
Authenticated encryption is clearly the superior approach in terms of security since it protects against spoofing, splicing, and even enables to implement protection against replay attacks~\cite{DBLP:journals/tcos/ElbazCGLPT09}.
However, the increase in security typically comes at the cost of increased latency and memory overhead to store the integrity information.

Regarding granularity and key handling, different approaches have been proposed so far.
Initial approaches relied on a single key for the whole encrypted memory (\eg Intel's TME, AMD's SME).
More recent designs, in contrast, also grant finer control over the used keys.
AMD's Secure Encrypted Virtualization (SEV), \eg supports the use of different keys for the virtualized guest machines.
Intel's Multi-Key TME (MKTME) even supports different keys with page-wise granularity by embedding the key ID directly into the physical address.
Our approach is perfectly compatible with all these design choices for key handling, but orthogonally extends them with support for tags in the virtual address space.
The actual encryption key for each memory block is, subsequently, derived from the page or root key and the respective tag.
As demonstrated in Section~\ref{sec:implementation}, even encryption with sub-cache line granularity can be supported in this way.

\section{Threat Model}
\label{sec:adversary_model_design_goals}
Based on the \tagenc architecture, we propose a hardware-assisted memory safety concept.
Similar to other threat models in the context of memory safety, we are considering an adversary using an exploitable memory bug to craft a memory vulnerability.
Based on the capabilities of the underlying memory encryption engine (encryption only or with authentication), our memory coloring design \tagenc provides two different levels of security guarantees.
\paragraph{\tauth Encryption \& Authentication:}
When detecting a spatial memory safety violation, \tagenc immediately triggers a system exception via the inbuilt authentication mechanism of the transparent memory encryption scheme.
Here, \tagenc is capable of detecting out-of-bound reads or writes, \ie a spatial memory bugs.
Furthermore, \tagenc also is capable of reporting the exploitation of temporal memory bugs, \eg use-after-free vulnerabilities.
\paragraph{\tenc Encryption:}
Compared to \tauth, this security policy limits the exploitation of spatial and temporal memory bugs.
For out-of-bound memory reads, \tagenc guarantees the confidentiality of the data stored in the target buffer.
Since the underlying memory encryption engine does not provide data integrity, \tagenc cannot maintain the integrity of data in the target buffer in an out-of-bound memory write.
However, \tagenc  with \tenc aggravates the exploitation of temporal bugs and spatial out-of-bound writes.

\section{Design}
\label{sec:design}
\begin{figure}[t]
  \begin{center}
    \includegraphics[width=0.95\linewidth]{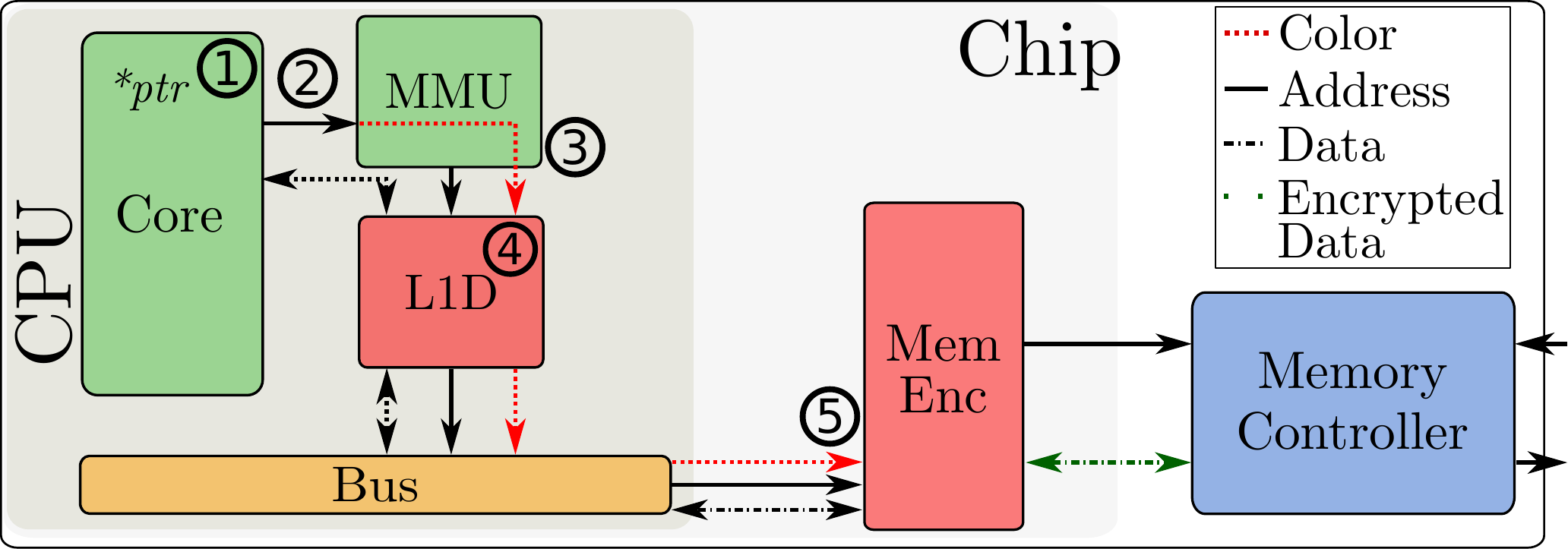}
    %\vspace*{-0.4cm}
    \caption{Overall \tagenc architecture. The memory encryption unit, placed between the memory subsystem and the memory controller, uses the color as a tweak.}
    %\vspace*{-0.6cm}
    \label{fig:lowrisc}
  \end{center}
\end{figure}

In our architecture, memory is allocated in software~\circleds{1} and a dedicated instruction assigns a random color to the memory object and stores it in the upper bits of the pointer~\circleds{2}.
When writing data to the memory, the color information is propagated through the MMU~\circleds{3}, the cache~\circleds{4}, and then finally is used as a tweak in the memory encryption unit~\circleds{5}.
On a memory access, the hardware transparently performs a cryptographic check without any further instrumentation.
This hardware architecture shown in Figure~\ref{fig:lowrisc} allows \tagenc to protect dynamic, local, and global data.

\subsection{Hardware}
\tagenc utilizes a built-in memory encryption unit to implement an efficient memory coloring scheme.
Initiated by a custom instruction, a random color is assigned for each memory object, which is stored in the upper unused bits of the pointer.
When accessing the colored memory object, it requires the correct color to be in place for the memory request.
\tagenc implements this \textit{lock-and-key} approach by using the color of the memory object as an additional input for the transparent memory encryption scheme.
Due to this strategy, each memory object colored with a random color is encrypted differently.

\subsubsection{Memory Coloring}
In \tagenc, the color of the memory object is assigned to the pointer.
Since memory allocations are a frequent task and assigning and generating a color in software is costly, a custom instruction using a hardware-based random number generator is used.
Similar to other designs~\cite{DBLP:conf/ccs/KwonDSKD13,arm2017pointerauthentication, DBLP:conf/acsac/SchillingWNM18}, \tagenc uses the upper bits of the pointer to store the color information.
This approach causes zero costs in terms of storing the color information and also minimizes any overhead to use pointers in software.
Since the address information and the corresponding color are already stored in the same register, there is no need to extensively modify the instruction set.
However, storing the color directly inside the pointer results in two disadvantages.
First, the number of colors, which influences the security of memory coloring, directly corresponds to the number of free bits in the pointer.
Second, using the upper bits of the pointer reduces the virtual address space of the system.
Nevertheless, in practice, a trade-off between the available address space and security guarantees can be found.
In most 64-bit platforms, already a reduced address space with free upper bits is used.
For example, the AArch64 Linux port~\cite{LinuxArm20} limits, by default, the virtual address space to \SI{39}{\bits} and, therefore, supports colors up to \SI{25}{\bits}.
While this address space might be sufficient for, \eg mobile devices, a \SI{512}{\giga\byte} address space is not acceptable for high-performance servers.
By using the larger \SI{48}{\bits} addressing model, the address space can be extended to address \SI{256}{\tera\byte} and supporting colors up to \SI{16}{\bits}.
Security limitations of different color sizes are discussed in Section~\ref{sec:security}.
When using a colored pointer, the color needs to be propagated throughout the system up to the memory encryption.
Since the MMU of the processor only considers the lower bits of a pointer to translate the virtual to the physical address, \tagenc needs to bypass the MMU translation and directly forwards the color information to the cache (see Figure~\ref{fig:lowrisc}).

\paragraph*{Cache Architecture.}

\begin{figure}[t]
  \begin{center}
    \includegraphics[width=1\linewidth]{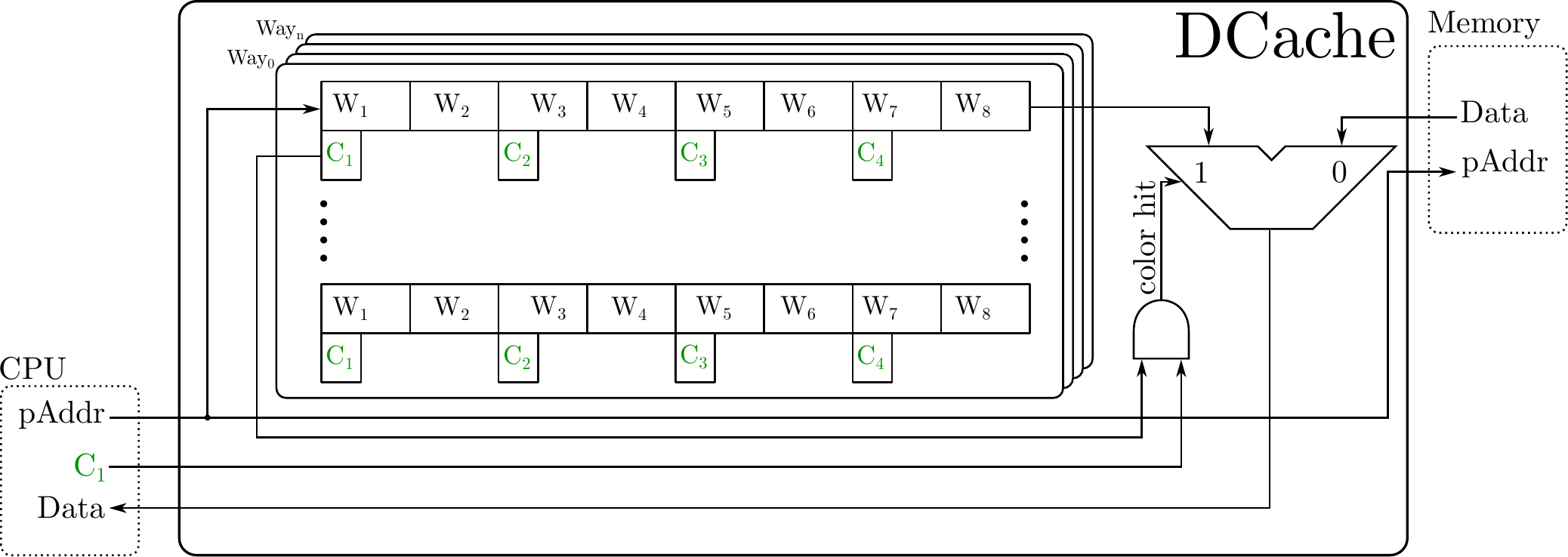}
    %\vspace*{-0.5cm}
    \caption{Set-associative cache architecture extended to support a color for each \textit{TG}-bytes. On a cache hit, the data cache also checks the color. On a tag mismatch, the cache line is fetched from the memory.}
    %\vspace*{-0.65cm}
    \label{fig:cache}
  \end{center}
\end{figure}

Figure~\ref{fig:cache} shows our extension to the cache architecture.
In \tagenc, each $TG$-bytes of memory $W$ are tagged with a $TS$-bits color $C$.
For cache management, the color $C$ is also stored in the cache for each memory object $W$.
A cache hit is only valid if the color stored in the cache matches the color stored in the pointer.
The design of \tagenc also supports sub-cache line tag granularities, \eg one color for two words, which can be configured.
In Section~\ref{sec:cache}, we present a concrete cache implementation supporting the proposed color management.

\subsubsection{Memory Encryption}
When writing to memory, the color is used as a tweak to encrypt the data using the transparent memory encryption unit.
To decrypt this data on a memory access, the read operation needs to have the correct color stored in the upper bits of the pointer.
Depending on whether the memory encryption unit provides encryption and authentication or solely encryption, \tagenc either implements detection strategy \tauth or \tenc.

\paragraph{\tauth: Exception-based Notification.}

This strategy is possible for memory encryption engines~(MEE) providing encryption and authentication.
The exception-based notification policy immediately triggers an exception if the system performs a wrong memory access on a color mismatch.
The encryption operation $C, T = Enc_{AE}(k, t, P)$ takes the encryption key $k$, the color as the cipher tweak $t$, and the plaintext data $P$ as an input to compute the ciphertext $C$ and the authentication tag $T$ as the output.
Both the ciphertext and the authentication tag are stored inside the memory, while the color is \textit{not}.
When reading data back from memory, the MEE verifies the integrity of the ciphertext and decrypts the data, \ie $P \parallel \bot = Dec_{AE}(k, t, C, T)$.
On a successful ciphertext verification using the authentication tag, the cipher returns the correct plaintext data $P$.
If the integrity verification fails, \eg owing to a wrong color, the MEE returns an error $\bot$ and the system automatically triggers an exception indicating an invalid memory access.

\paragraph{\tenc: Detection-based Notification.}

The detection-based notification policy corrupts the data when performing a wrong memory access.
Here, the architecture uses a tweakable block cipher $C = Enc(k, t, P)$ without authentication.
The ciphertext $C$ is computed using the encryption key $k$, the color as the tweak $t$, and the plaintext data $P$.
When reading data from memory, the MEE automatically decrypts the ciphertext under the encryption key and the tweak given by the memory color stored in the address.
On a correct memory access, this also returns the correct plaintext data.
However, using the wrong color on a malicious memory access decrypts the ciphertext with the wrong tweak leading to an invalid plaintext.

\subsection{Software}
\label{sec:software}
The \tagenc hardware architecture alone does not thwart memory safety errors.
It requires software support and protecting all memory allocations to detect most spatial and temporal memory vulnerabilities.
The principle idea of the memory protection is that all memory allocations are colored, meaning that every associated pointer to a memory allocation stores a color in the upper bits of the pointer.
Only when using the pointer with the correct color, the memory access is successful.
Otherwise, depending on the detection strategy, either an error is raised or the payload data is destroyed.
In this section, we describe how to use the hardare design in software to thwart memory safety vulnerabilities.

\paragraph{Heap Data.}

On each dynamic memory allocation, \eg via a call to \texttt{malloc}, the returned pointer is assigned a dedicated random color.
Furthermore, the memory is properly aligned to match the tag granularity $TG$.
As discussed previously, our design uses a dedicated hardware instruction to perform this operation and, therefore, only adds a small overhead to manage the colors.
When accessing heap data later on, every access encrypts or decrypts the data automatically using the assigned tag information.
When releasing dynamic memory through a \texttt{free} operation, the color information of the pointer is removed and the memory is released to the OS.

\paragraph{Local Data.}

Local allocations on the stack are aligned to match the tag granularity $TG$.
The corresponding pointer is colored using the custom instruction.
Further accesses then encrypt or decrypt the data when accessing the memory.

\paragraph{Global Data.}

Protecting global data requires more effort.
There are two possibilities to deal with global data.
First, the protection of global data can be realized during compile time, where the compiler assigns each global variable a dedicated color.
During the compilation, initialized global data is then encrypted using the pre-assigned color so that memory accesses in the program yield the correct data.
However, this approach requires additional overhead to manage the colored pointers in software.
Furthermore, access to global data always uses the same color for encryption, enabling \eg replay attacks.
To avoid the problem of replay attacks and unnecessary color management in software, we aim for a second approach.
Our design replaces all references to global variables with a new pointer.
Additionally, the compiler adds a dedicated startup hook function for each global variable.
During the startup of the program, this hook function first colors the new pointer.
Second, it reads the unencrypted global data from the executable and then writes this data to memory using the new colored pointer.
Thereby, the global data automatically gets encrypted using the colored pointer.
This approach allows us to use the same instruction to randomly color the new pointer.
By using a random color in the pointer, we also mitigate replay attacks since the global data is encrypted differently at every program start.
While this approach enhances security, it also simplifies the software support.

\section{Implementation}
\label{sec:implementation}
In this section, we introduce the base platform where we integrate \tagenc and show the necessary hardware extensions.
We discuss the color generation and propagation and further explain how the memory encryption framework is used to implement the coloring scheme.
Finally, we introduce the compiler extension utilizing the \tagenc architecture to protect data.

\paragraph{Base Platform.}

We build the prototype for \tagenc on top of the CVA6 platform~\cite{DBLP:journals/tvlsi/ZarubaB19}, a system-on-chip~(SoC) using a 64-bit 6-stage RISC-V processor supporting to run Linux when mapped to a FPGA.
The CVA6 is extended with the open-source memory encryption scheme \memsec~\cite{DBLP:conf/fpl/WernerUSSM17}.
\memsec is placed between the data cache and the DDR3 memory controller and automatically encrypts all data leaving the processor.
Furthermore, we adapted the data cache and increased the cache line size from 16 to 64 byte.

\subsection{Hardware Extensions}
The necessary hardware extensions to implement \tagenc are minimal and only require two adaptions.
First, the system requires a mechanism to create a color and to propagate it through the system.
Second, the memory encryption engine~(MEE) needs to be extended to handle the additional color input for the cipher.

\subsubsection{Color Generation}
\label{sec:tag_generation}
Tagging a memory region with a dedicated memory color is initiated in software.
Thus, we extend the \riscv instruction set with a dedicated instruction to allow performing this operation efficiently in software.

\paragraph{\texttt{mstp rd,rs}.}
To color a pointer, the custom instruction \mstp is added to the \riscv instruction set.
This instruction takes the value from the source register $rs$ (typically the pointer), colors it, and stores the result to the destination register $rd$.
Our architecture uses the SV39 addressing model~\cite{NON:waterman_riscvpriv191_2016} of \riscv, where the lower \SI{39}{\bits} of the virtual address space are used.
The remaining upper bits of the pointer are set to the sign bit of the pointer value (either all-zero or all-one).
The \mstp instruction colors the pointer and replaces the upper \SI{25}{\bits} ($TS$-bits) with a random color value.
To differentiate between a colored and a non-colored pointer, the color bits cannot be set to all-zero or all-one.
The random color value is generated using a hardware-internal pseudo-random number generator, which is initialized during processor startup with a software inaccessible seed value.

\subsubsection{Color Propagation}
After instrumenting a pointer with the color bits, the MMU translates the virtual to a physical address.
In SV39 of \riscv, the MMU only uses the lower \SI{39}{\bits} of the address for its translation.
The upper \SI{25}{\bits} containing the color information bypass the MMU's address translation.
Both, the physical address and the color bits, get processed by the L1 data cache and are then propagated to the MEE via the processor's bus architecture.

\subsubsection{Cache Design}
\label{sec:cache}
As data in \tagenc is tagged with the color of the corresponding pointer, the cache also needs to be aware of these colors.

\paragraph{Colors.}
The prototype implementation of \tagenc uses a tag granularity~($TG$) of 16-bytes and a color size~($TS$) of \SI{25}{\bits}.
As denoted in Figure~\ref{fig:cache}, each 64-byte cache line stores four $TS$-bits colors.
Internally, the cache differentiates between three values for a color: \textit{no color}, \textit{valid color}, and \textit{invalid color}.
When accessing the cache with an address where there is no color stored inside (the upper bits are all-zero or all-one), the cache-internal color is set to \textit{no color}.
A \textit{invalid color} color is stored in the cache when prefetching a cache line with the wrong color triggers a decryption exception.
When accessing the cache with an instrumented pointer, the \textit{valid color} information is stored in the cache.

\paragraph{Cache Hit.}
A cache hit is triggered when having the correct data \textit{and} correct color in the cache.
If there is a color mismatch, a cache miss is triggered.

\paragraph{Cache Miss.}
On a cache miss, the cache architecture issues a memory read request to the main memory.
Here, the colors of the cache line are used as a cipher tweak to decrypt data from the memory.
After fetching the decrypted data from memory, the colors are stored in the metadata structure of the cache.
In detection strategy \tauth, accessing the cache with a wrong color leads to a decryption and verification error and the error is forwarded to the processor as an exception.

\paragraph{Cache Prefetching.}
Cache prefetching is used to reduce the latency for memory accesses by precautionary fetching an entire cache line from memory.
This technique speeds up memory accesses but also challenges our colored cache architecture.
When performing a cache prefetch, only the color of the first $TG$-bytes of the cache line is known.
To decrypt the remaining cache line, the system assumes the later colors of the cache line are the same and uses the first color to decrypt the whole cache line.
Due to the memory fragmentation, this assumption is correct with high probability and the MEE decrypts the cache line correctly.
Furthermore, the used color is copied to all color entries of the cache line.
In case of a wrong decryption operation due to prefetching, detection strategy \tauth detects a wrong decryption and thus invalidates the color entry in the cache but does not raise an exception.

\paragraph{Cache Eviction.}
During cache eviction, a dirty cache line is written back to memory and is encrypted using the colors stored inside the cache.
In case of having invalid colors in the cache, \ie due to prefetching, the cache only issues memory updates for entries with valid colors.
Invalid cache entries are filtered and not written back to memory.

\subsubsection{Memory Encryption Engine}
\memsec, which is directly placed between the cache and the memory controller, transparently encrypts all data leaving the processor on bus level.
Similar to other MEEs, the encryption key is randomly generated during the device startup.
For the \tagenc architecture, we extended the MEE to support the additional color input.
To implement detection strategy \tauth, we use the authenticated encryption cipher ASCON~\cite{Dobraunig2016}, which provides data confidentiality and integrity.
We tweak the cipher by using the size-extended color value for the nonce input of the cipher.
To re-initialize already encrypted memory, \memsec also allows to suppress authentication errors using a defined memory pattern.
For implementing strategy \tenc, we use the low-latency tweakable block cipher QARMA~\cite{DBLP:journals/iacr/Avanzi16}, which is also used in ARM's pointer authentication scheme~\cite{arm2017pointerauthentication}.
Since QARMA natively supports an additional input, we use the size-extended color value as input for the tweak.

\subsection{Software Extensions}
\label{sec:implemetation_software}
To detect memory safety violations and to protect every memory allocation, we need software support.
We extended an LLVM-based C compiler~\cite{DBLP:conf/cgo/LattnerA04} with an LLVM IR pass and a tiny runtime support library.
The compiler needs to protect three storage classes: heap, local, and global data.

\paragraph{Protection of Heap Data.}

Protection of heap data is accomplished by using the GNU linker functionality to create wrappers around heap functions such as \texttt{malloc}, \texttt{free}, and \texttt{realloc}.
For policy \tenc, the \texttt{malloc} wrapper aligns the size argument to $TG$ and then calls \texttt{malloc} itself.
Because heap allocation for 64-bit RISC-V systems is already \SI{16}{\byte} aligned, we only have to apply the \texttt{mstp} instruction on the pointer before returning to the application.
Hence, the overhead therefore is negligible.
In Listing~\ref{lst:wrap_malloc} we show the implementation of the wrapped \texttt{malloc} function for \tenc of the runtime library.
When utilizing \tagenc for detection policy \tauth, \texttt{malloc} additionally initializes the memory with its color.
Since this memory object could already be encrypted with a different color, na\"ively re-coloring would trigger an authentication error.
Thus, \tagenc uses the nullification mechanism of \memsec to initialize the memory.

% \notemw{maybe use figure instead of listing. (differences in caption styling)}
% \begin{listing}[t]
%   \begin{minted}{c}
%     void* __wrap_malloc(size_t size) {
%       size = roundup(size);
%       void *ptr = __real_malloc(size);
%       if (ptr == NULL) return NULL;
%       return mstp(ptr);
%     }
%   \end{minted}
%   \caption{Malloc wrapper that tags the returned pointer in \tenc.}
%   \label{lst:wrap_malloc}
% \end{listing}

% Reset Counter for figure being used for listings. We want to start with listing 1
\renewcommand{\figurename}{Listing}
\newcommand{\oldcounter}{\value{figure}}
\setcounter{figure}{0}
\begin{figure}[H]
  \begin{center}
    \includegraphics[width=0.95\linewidth]{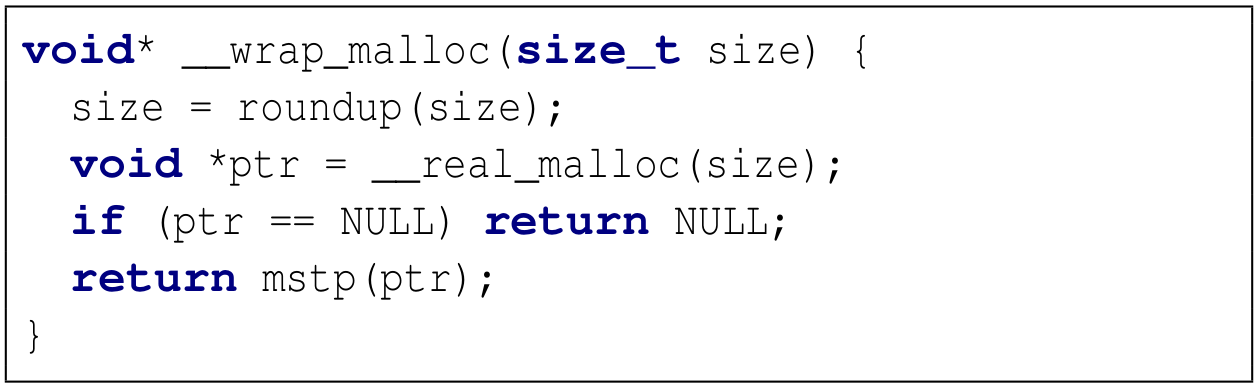}
    %\vspace*{-0.4cm}
    \caption{\tenc \texttt{malloc} wrapper tagging the returned pointer.}
    %\vspace*{-0.4cm}
    \label{lst:wrap_malloc}
  \end{center}
\end{figure}
% Reset counter back to the original value
\setcounter{figure}{3}
\renewcommand{\figurename}{Figure}
For \texttt{free}, the wrapper removes the color from the pointer argument and then calls the \texttt{free} function.
This operation is done purely in software by using two shift operations.
Notice that the heap administration data is stored in plain in between the encrypted heap data.
Writing via a heap data-pointer out of bounds into the heap administration overwrites plain data with encrypted data making it hard for an attacker to modify the heap administration in a controlled way.

\paragraph{Protection of Local Data.}

Local variables on the stack are protected by scanning for \texttt{AllocaInst} instructions through a custom LLVM IR compiler pass.
For each \texttt{AllocaInst}, we align the size argument to $TG$ bytes, we align its address alignment to $TG$, and we insert an \mstp instruction between the \texttt{AllocaInst} instructions and all its users.
Additionally, for \tauth, we re-initialize the allocated memory with the assigned color.
We perform a simple analysis on \texttt{AllocaInst} instructions to exclude protection of cases where incorrect usage will not be possible.
For example, cases where the result of \texttt{AllocaInst} is not used by a \texttt{GetElementPtrInst} instruction with non-constant indices and the result is not stored in memory or leaves the function as argument of a function call.

\paragraph{Protection of Global Data.}

For each global definition/declaration of a variable named \texttt{foo}, we create a new global definition/declaration of a pointer called \texttt{\_\_foo\_mst} that points to \texttt{foo}.
Furthermore, the LLVM IR pass replaces all references to \texttt{foo} with \texttt{\_\_foo\_mst} to get a colored pointer to \texttt{foo}.
The runtime support library is informed about definitions \texttt{foo}, the size of \texttt{foo}, and the new pointer \texttt{\_\_foo\_mst} via a constructor function.
During the startup, the constructor function in the runtime library will insert a color on \texttt{\_\_foo\_mst} by means of an \mstp instruction.
It will also encrypt \texttt{foo} using the color that has been put on \texttt{\_\_foo\_mst}.
This happens by simply reading the data in plain using the original all-zero pointer and then writing it back to memory using \texttt{\_\_foo\_mst}.
The runtime overhead of referencing a global variable is therefore one-load instruction.
Furthermore, notice that on every execution of a protected application, its global variables will get different colors.

As with local data, we perform an analysis to exclude protection of (static) global data where we can prove that incorrect usage is not possible.
Protecting global variables that have initializers with pointers to global variables complicates the protection.
The runtime support library is informed via a constructor function about the positions of these pointers in global the initializers and patches those pointers with the color of the global variable it is pointing to.

\paragraph{Backward Compatibility.}

Application code that is protected by \tagenc can be combined with unprotected code.
For example, the unprotected pointer results of \texttt{fopen()} or \texttt{mmap()} can be used in protected code without problems.
Furthermore, protected pointers of heap, local, or global data can arbitrarily be passed to unprotected library functions without problems.
The only compatibility issue that we are aware of is sharing global variables between protected and unprotected code, \ie \texttt{stdout}.
Unprotected code will expect it unencrypted, while protected code will expect the data to be encrypted.
Due to indirection via \texttt{\_\_foo\_mst} this issue will result in linking errors rather than runtime errors.
The user should then manually place these global variables on a blacklist of global variables that are not suitable for protection.

\paragraph{Pointer Arithmetic.}
Incrementing a pointer to reach data within a colored object is natively supported in the \tagenc architecture because the color information in the upper bits of the pointer is not altered.
However, subtracting or adding two pointers or performing shifts or multiplications on such pointers can modify the color and is therefore dangerous.
To also support these operations and enable safe arithmetic operations avoiding integer overflows on colored pointers, dedicated instructions could be added.
In the Armv8.5-A instruction set~\cite{ARM82020} supporting MTE, dedicated add and subtract instructions, ignoring the upper bits, are used for pointer arithmetic.

\section{Performance Evaluation}
\label{sec:evaluation}
To quantify the hardware and software overhead, we synthesize the \tagenc architecture for a Xilinx Kintex-7 series FPGA and run a recent Linux operating system on our platform.
We report the performance and hardware overhead introduced by \tagenc for the $TS$=25 and $TG$=16 memory coloring configuration using different applications, from microbenchmarks to application code.

\subsection{Hardware Overhead}
In Table~\ref{tab:hwoverhead}, we show the hardware overhead for the FPGA design in terms of lookup-tables~(LUTs) and flip flops~(FFs).
The overhead numbers include the changes required for the new instruction, the color propagation, and the extended cache.
Clearly, the hardware overhead of less than 1\,\% is very attractive and negligible in practice.

\begin{table}[b]
  \centering
  %\vspace*{-0.3cm}
  \caption{Hardware overhead for the $TS$=25 and $TG$=16 memory coloring configuration.}
  %\vspace*{-0.3cm}
  \label{tab:hwoverhead}
  \begin{tabular}{lllll}
    \toprule
    \multirow{3}{*}{Config.} & \multicolumn{2}{c}{LUTs} & \multicolumn{2}{c}{FFs} \\
                             & Baseline & Overhead & Baseline & Overhead \\
                             & [LUTs]   & [\%]     & [FFs]    & [\%] \\
    \midrule
    ASCON                    & 57386 & 0.53  & 33885 & 0.14 \\
    QARMA                    & 55804 & 0.67  & 32173 & 0.18 \\
    \bottomrule
  \end{tabular}
\end{table}

\paragraph{Cache Architecture.}
For \tagenc, the cache architecture is extended to also store colors along with the data.
Furthermore, the decision logic to detect cache hits and misses is extended to also consider the colors in the cache.
The hardware overhead for this comparison logic is relatively small compared to the overhead for storing the colors.
The required hardware overhead for storing the colors in the cache is a function of the used color size $TS$ and tag granularity $TG$.
Note, our design only needs to store colors in the cache and there is no need for a separate, large cache for colors as it is required in other architectures~\cite{DBLP:conf/iccd/JoannouWKMBXWCR17,song2015towards} to speed up accessing tags in memory.
Thus, the design not only has less hardware overhead but also improves the runtime latency and bandwidth, as there are no memory accesses for colors needed.
For an $m$-way $n$-set associative cache with a cache line size of $C$ bytes, a tag granularity of $TG$ bytes, and a color size of $TS$ bits, we can compute the required number of color bits $T$ as defined in Equation~\ref{eq:overhead:tag_overhead}.

\begin{equation}
  T = \text{nSets } \cdot \text{mWays } \cdot TS \cdot \frac{C}{TG}
  \label{eq:overhead:tag_overhead}
\end{equation}

Our modified \SI{16}{\kilo\byte} data cache of the CVA6 core is organized as a 4-way 64-set associative cache with a 64 byte cache line.
For a tag granularity of $TG$=64 byte and a color size $TS$=\SI{8}{\bits}, the overhead for storing the colors is \SI{2.05}{\kilo\bit}.
Although this is already fine-grained, our design even supports sub-cache line tag granularities, \eg $TG$=\SI{16}{\byte}.
For a configuration with $TG$=16 byte and $TS$=\SI{25}{\bits} the overhead for storing the colors is \SI{25.2}{\kilo\bit}.
Table~\ref{tab:cacheoverhead} shows the total number of color bits for different configurations including the corresponding overhead.

\begin{table}[b]
\centering
\caption{Number of additional cache bits and total cache overhead for storing colors.}
%\vspace*{-0.4cm}
\label{tab:cacheoverhead}
\begin{tabular}{lll}
\toprule
Cache Configuration & \begin{tabular}[c]{@{}l@{}}Color bits T\\ {[}bit{]}\end{tabular} & \begin{tabular}[c]{@{}l@{}}Cache Overhead\\ {[}\%{]}\end{tabular} \\ 
\midrule
TS=8, TG=64         & 2048                                                             & 1.56                                                              \\
TS=8, TG=16         & 8192                                                             & 6.25                                                              \\
TS=25, TG=64        & 6400                                                             & 4.88                                                              \\
TS=25, TG=16        & 25600                                                            & 19.53                                                             \\ 
\bottomrule
\end{tabular}
\end{table}

For our Xilinx-based FPGA, the cache is mapped to multiple block RAM instances.
For a small memory color configuration (course tag granularity and small color size), the color bits even fit in the already instantiated block RAM resources of the cache and thus have no impact on the block RAM utilization.
Only for the worst-case memory color configuration ($TG$=16 byte and $TS$=\SI{25}{\bits}), an additional block RAM module needs to be instantiated.
For other hardware technologies, \eg ASIC designs, the cache overhead directly results from the color size and tag granularity.
For, \eg a 64 byte cache line and a memory coloring configuration of $TG$=16 byte and $TS$=\SI{25}{\bits}, a cache line is extended by \SI{100}{\bits} for storing colors, resulting in an overhead for the cache of $19.5\,\%$.

\subsection{Runtime Overhead}
\label{sec:evaluation_runtime}
To measure the software overhead of our system, we compiled different binaries with our custom LLVM-based toolchain protecting all dynamic memory, all locals, and all globals.
We evaluate the performance overhead by running different benchmark applications in user mode on the Linux environment running on our FPGA hardware implementation.
Note that the entire system, including the Linux operating system, is executed in the encrypted memory domain, but only user applications are additionally instrumented and use memory coloring.
We use a set of benchmarks, including SPEC 2017 and smaller microbenchmarks, such as SciMark2 and MiBench.
\tagenc, which we envision to be an extension of systems already featuring a memory encryption unit, adds an overhead between 1.5\,\% and 6.1\,\% on average for thwarting logical memory safety vulnerabilities on such a system.
As we do not have access to state-of-the-art commercial memory encryption engines, which typically yield a performance overhead between 5\,\% and 26\,\%~\cite{ARMMENC}, we further evaluate the performance of the open-source \memsec framework.
For transparently encrypting the whole external memory, we measured a performance overhead between 58.9\,\% and 109.8\,\% for SPEC 2017.
The overall combined overhead for thwarting physical and logical memory safety vulnerabilities with \tagenc and \memsec is between 62.0\,\% and 116.1\,\%.

\paragraph{Code Size Overhead.}
As we are linking the 244 bytes runtime library while building the binary, the code size overhead to protect dynamic memory is constant and negligible compared to the overall binary size.
Furthermore, to also protect local and global variables, \mstp instructions are inserted as part of an instrumentation pass.
Compiling the \enquote{SPECspeed 2017 Integer} testsuite with our toolchain, adds an average code size overhead of 1.02\,\%.

\paragraph{Runtime Overhead of \tagenc.}
Instrumenting pointers with the \mstp instruction and aligning memory objects to the tag granularity increases the number of instructions for executing a program.
Furthermore, prefetching data can possibly lead to invalid colors and, thus, requires additional memory accesses slowing down the execution.
However, such cases only occur rarely and are not relevant in practice.

\paragraph{SPEC CPU 2017}
To quantify the performance overhead introduced by our architecture, we first measured the performance overhead introduced by the memory encryption framework.
Then, we analyzed the overhead additionally introduced by the memory coloring scheme.
For our evaluation, we used a subset of C/C++ benchmark programs of the \enquote{SPECspeed 2017 Integer} and \enquote{SPECrate 2017 Floating Point} testsuites.
However, due to the missing OpenMP support for \riscv in LLVM, programs that depend on that are omitted.
Furthermore, due to the constrained hardware resources of the FPGA prototype, several benchmarks already failed with an out-of-memory exception (\texttt{mcf}, \texttt{omnetpp}, \texttt{x264}, and \texttt{lbm}) and one with a runtime exception (\texttt{xalancbmk}).
Note, these benchmarks already failed on the unmodified base platform with the native LLVM-compiler.
Similar to other memory safety tools~\cite{DBLP:conf/cc/DuckY16, ASAN2020}, \tagenc also found several memory bugs in \texttt{perlbench}.

\begin{figure}[t]
  \begin{center}
    \includegraphics[height=3.5cm]{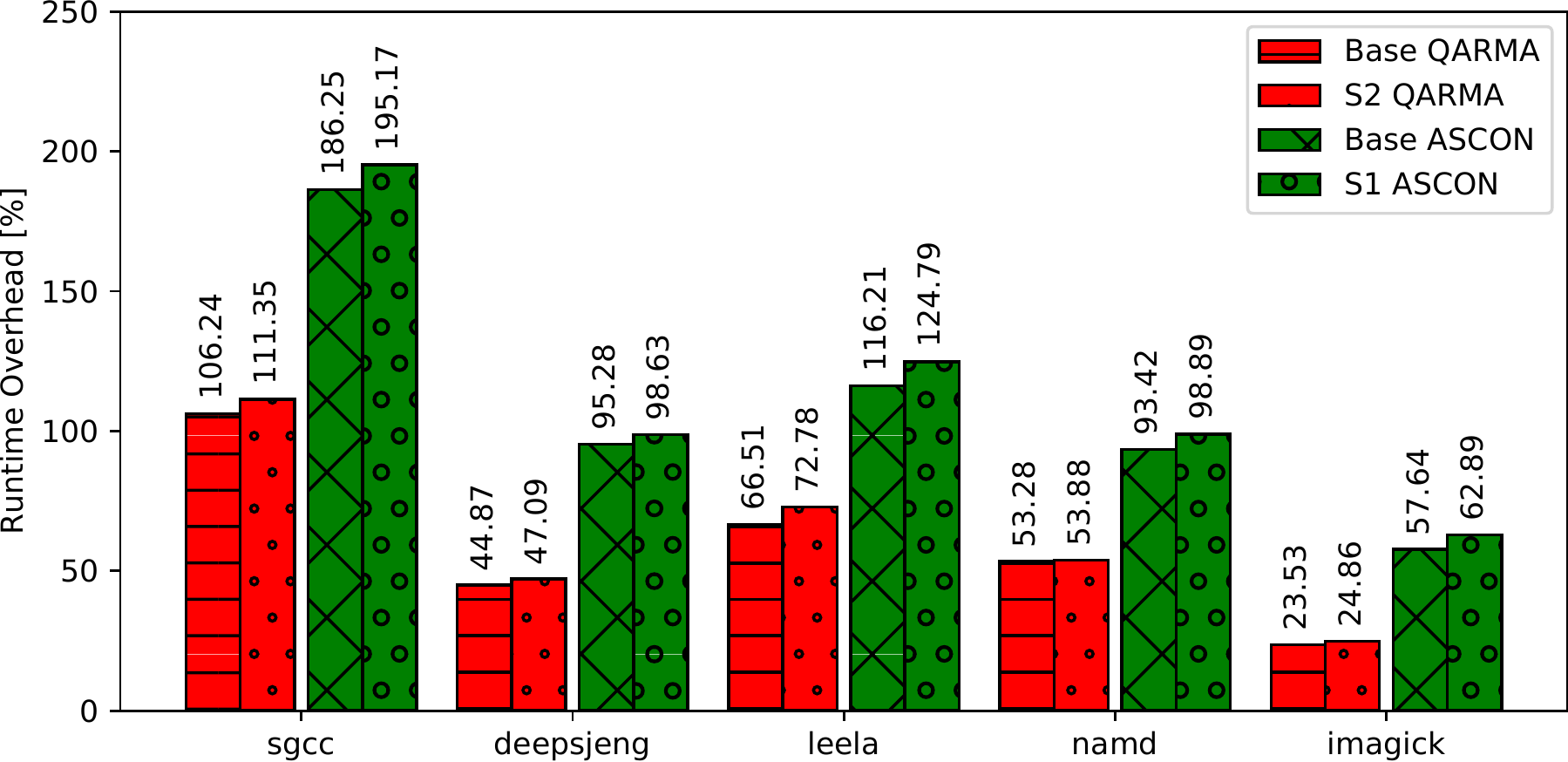}
    %\vspace*{-0.35cm}
    \caption{Runtime overhead for SPEC CPU 2017.}
    %\vspace*{-0.5cm}
    \label{fig:runtime_spec}
  \end{center}
\end{figure}

The performance overheads for the subset of \enquote{SPECspeed 2017 Integer} and \enquote{SPECrate 2017 Floating Point} benchmarks are depicted in Figure~\ref{fig:runtime_spec}.
This diagram shows that the dominating performance factor is the used memory encryption engine and \emph{not} \tagenc.
The relative overhead for thwarting logical memory safety vulnerabilities on a system already featuring memory encryption is 5.2\,\% for \tenc and 6.1\,\% for \tauth on average.
For the unoptimized \memsec framework, we measured a performance overhead of 58.9\,\% for QARMA and 109.8\,\% for ASCON on average compared to the base platform.
The overall performance overhead for the combined physical and logical memory safety protection averages to 62.0\,\% for \tenc and 116.1\,\% for \tauth.

\paragraph{SciMark2.}

\begin{figure}[t]
  \begin{center}
    \includegraphics[width=\linewidth]{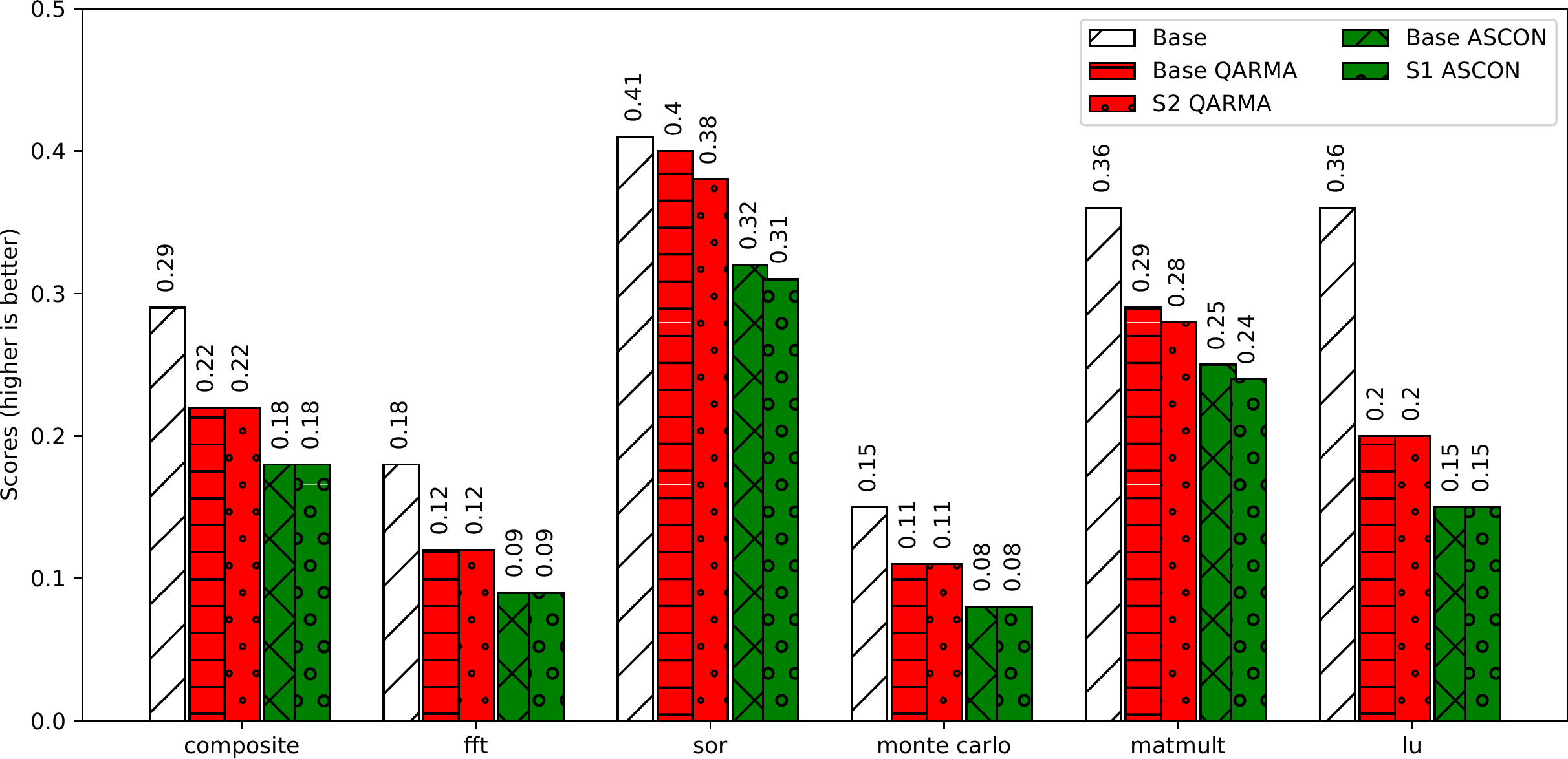}
    \caption{Benchmark score for SciMark2.}
    \label{fig:resultscimark}
  \end{center}
\end{figure}

To evaluate realistic computing workloads, we use SciMark2~\cite{pozo98scimark}, a benchmark for scientific and numerical computing.
We both measure the benchmark performance using SciMark2's internal test score, which is shown in Figure~\ref{fig:resultscimark}, and the runtime overhead using the hardware prototype.
Again, the relative overhead for logical memory safety protection is with 3.9\,\% for \tenc and 4.79\,\% for \tauth low.
When comparing the CVA6 base platform to the system featuring \memsec as MEE, we measured an average runtime overhead of 55.45\,\% for QARMA and 69.09\,\% for ASCON.
For the combined protection, we determined a performance overhead of 61.51\,\% for \tenc and 77.19\,\% for \tauth on average when compared to the system without the MEE.

\paragraph{MiBench.}

\begin{figure}[t]
  \begin{center}
    \includegraphics[width=\linewidth]{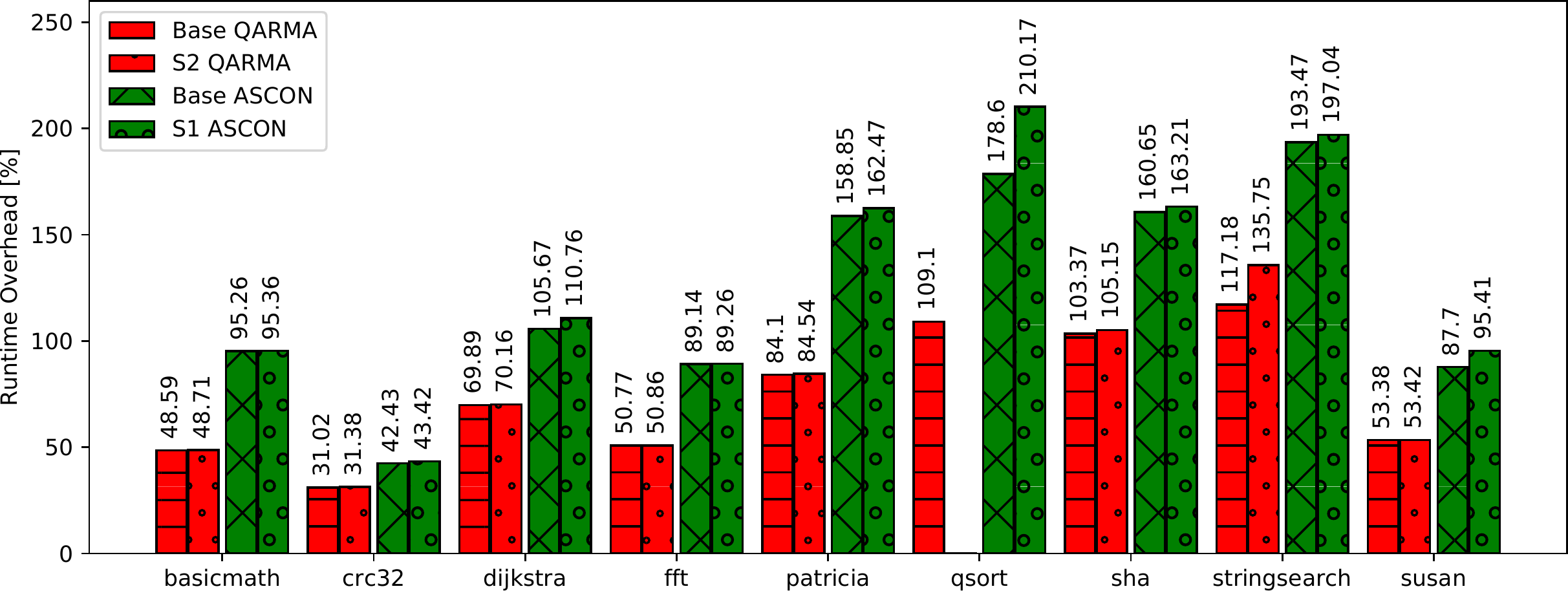}
    \caption{Runtime overhead for MiBench.}
    \label{fig:cycles_mibench}
  \end{center}
\end{figure}

MiBench~\cite{guthaus2001mibench} is a free, representative, embedded benchmark suite and it is used by us to evaluate the runtime overhead of our design on application-level code.
In Figure~\ref{fig:cycles_mibench}, we show the runtime overhead in terms of processor cycles relative to the baseline without memory encryption.
The evaluation was performed using 10,000 test runs to average out scheduling effects from the operating system.
Protecting all dynamic memory, all locals, and all globals with \tagenc introduces an average overhead of 1.5\,\% for \tenc and 4.9\,\% for \tauth on the system already featuring memory encryption.
Our measurement shows that \memsec adds an average overhead of 74.2\,\% for QARMA and 123.5\,\% for ASCON compared to the baseline without MEE.
For thwarting physical and logical memory safety vulnerabilities, we measured a combined overhead of 76.6\,\% for \tenc and 129.7\,\% for \tauth on average.

\subsection{Prototype Limitations}
\label{sec:plim}
As seen in Section~\ref{sec:evaluation_runtime}, the main factor of the runtime overhead is the MEE and \emph{not} \tagenc.
Hence, our performance evaluation largely is affected by the performance of the underlying memory encryption unit.
However, \memsec, the only, to the best of our knowledge, open-source MEE available, is not optimized for throughput and latency.
\begin{figure}[t]
  \begin{center}
    \includegraphics[width=\linewidth]{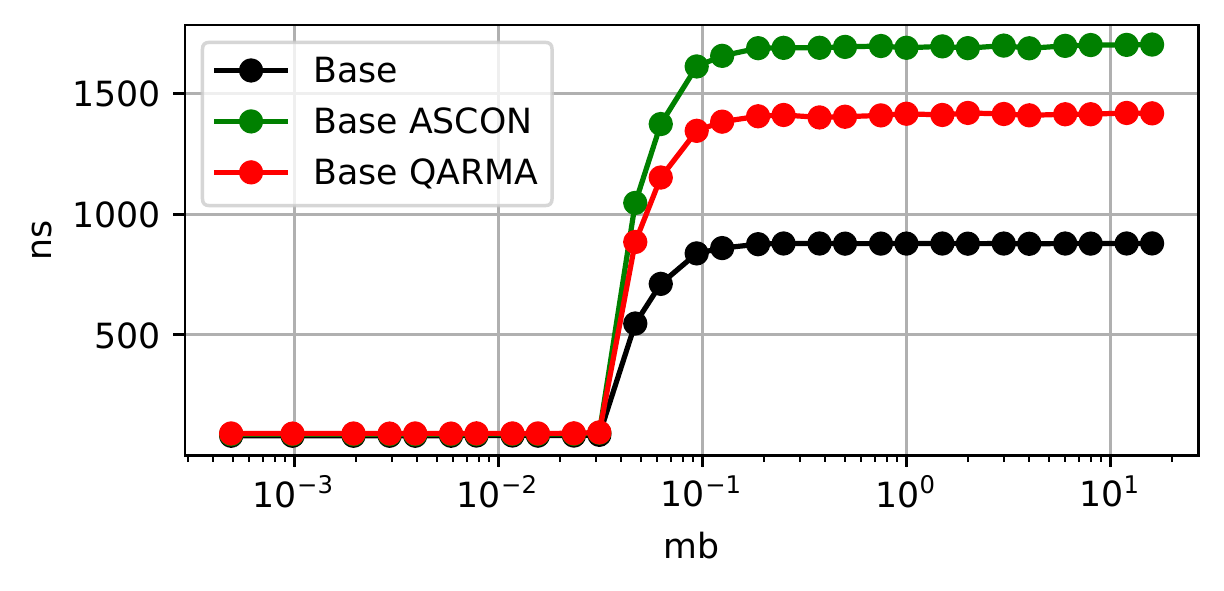}
    \caption{Memory latency measured with LMBench.}
    \label{fig:latency_lmbench}
  \end{center}
\end{figure}
Figure~\ref{fig:latency_lmbench} depicts the significant impact of \memsec with ASCON and QARMA on the latency measured by the \texttt{lat\_mem\_rd 64M 512} benchmark of LMBench~\cite{DBLP:conf/usenix/McVoyS96}.
The memory throughput, measured with \texttt{bw\_mem 4M rdwr}, also dropped from 52 MB/s to 14 MB/s for QARMA and 10.9 MB/s for ASCON.
In comparison, state-of-the-art encryption engines typically yield a performance penalty between 5\,\% and 26\,\%, as reported by ARM~\cite{ARMMENC}.
Although optimizing \memsec or designing a high-speed MEE is not part of our contribution, we point out different optimization strategies in Section~\ref{sec:conclusion}.
Finally, we want to emphasize that we envision \tagenc to be an extension of systems already featuring a transparent memory encryption scheme.
With major vendors, such as Intel with SGX, TME, and MKTME~\cite{ars} and AMD with SME and TSME~\cite{Kaplan2016a}, highlighting the importance of memory encryption, we expect an increasing number of such systems in the near future.
Here, \tagenc proposes an efficient solution to realize memory coloring on top of such systems with performance overheads between 1.5\,\% and 6.1\,\%.

\section{Security Evaluation}
\label{sec:security}

\tagenc enhances security guarantees of applications by mitigating the exploitation of most temporal or spatial memory bugs.
Based on the underlying MEE, \ie encryption and authentication or encryption only, \tagenc either enforces security policy \tauth or \tenc.

\paragraph{Spatial Memory Safety in \tauth.}
Spatial memory bugs allow an adversary to access data outside of the objects bound.
To detect these bugs, \tagenc utilizes the architecture to color the pointer and to initialize the memory object with a random color on a memory allocation.
Any subsequent access to this colored object requires that the access pointer is colored with the identical color, or an authentication error is triggered by the MEE.
Hence, out-of-bound read or write accesses to memory objects with a wrongly colored pointer are detected by \tagenc in \tauth.
Similar to other tagged memory schemes, \tagenc cannot detect intra-object overflows.
Since the vulnerable buffer, as well as the target, are stored in the same memory object, both objects have the same color.

\paragraph{Spatial Memory Safety in \tenc.}
Compared to \tauth, this security policy limits the exploitation of spatial memory bugs.
Usually, the attacker either uses spatial memory vulnerabilities to leak sensitive data or to modify control or non-control related data to craft ROP, DOP, or other attacks.
In out-of-bound read accesses, data encrypted with the original color is decrypted using the wrong color of this pointer.
Hence, \tagenc with \tenc maintains the confidentiality of data in spatial out-of-bound reads and provides protection from attacks such as Heartbleed~\cite{DBLP:conf/imc/DurumericKAHBLWABPP14}.
Since the underlying MEE does not provide data integrity, \tagenc cannot prevent an attacker from overwriting data in a target buffer using an out-of-bound write.
However, when reading this data using the corresponding pointer, pseudorandom values are retrieved.
Using this pseudorandom value as control-flow related data, \eg as a return address, most likely will cause an exception.
As the attacker cannot overwrite data in the target buffer in a controlled way, \tagenc raises the complexity for performing data-oriented attacks, such as DOP.

\paragraph{Temporal Memory Safety in \tauth.}
In a temporal memory safety violation, a memory object is accessed after it is deallocated.
Temporal memory safety violations are mostly exploited by use-after-free vulnerabilities, which \tagenc with \tauth can detect.
In this attack, a memory object gets deallocated and the space, later on, is used by a new object.
The attacker then can use the dangling pointer either to leak sensitive data or to tamper data, \eg a vtable pointer.
A similar concept is used by a double-free attack, where the adversary calls the memory deallocation functionality twice.
\tagenc with \tauth mitigates such attacks by assigning a new color on each memory allocation and initialization, reading or writing by using the dangling pointer colored with the previous color will trigger an exception.
Since the current implementation of \tagenc does not re-color the memory object on deallocation, a memory read or write to this memory region using the dangling pointer cannot be detected by \tagenc.
However, as soon as a new memory object is allocated and initialized on this region, it is tagged with a new color and accesses using the dangling pointer can be detected.
To prevent this behavior, \tagenc could be, similar to ARM MTE, extended to colorize memory objects with a new color on each deallocation.

\paragraph{Temporal Memory Safety in \tenc.}
Although \tagenc with \tenc cannot detect temporal memory bugs, it prevents the adversary from leaking data, \ie \tagenc maintains the confidentiality of data.
When using this vulnerability to overwrite sensitive data, such as vtable entries, the attacker cannot insert targeted data because the wrong color of the dangling pointer for the encryption is used.

\paragraph{Null Pointers.}
Pointers created in external libraries, which are not compiled with \tagenc, are not colored and thus have the all-null color.
Unlike CHERI~\cite{DBLP:conf/isca/WoodruffWCMADLNNR14}, where only the pointer is tagged with additional metadata and not the memory itself, \tagenc explicitly tags memory objects with its color.
Hence, a read or write vulnerability on a null-colored object only allows the attacker to access other null-colored memory objects and not the whole memory.
Colored data that is allocated by protected code can be passed to unprotected code, \eg external libraries, and is also protected there.

\paragraph{Stack Coloring.}
By coloring the stack pointer with \mstp on program initialization, all objects on the stack, which are not explicitly colored, \ie stack spills, are assigned a random color.
This strategy allows \tagenc to separate the stack from null-colored objects, \eg objects created in unprotected code.

\paragraph{Entropy.}
Similar to countermeasures like ASLR, PARTS~\cite{DBLP:conf/uss/LiljestrandNWPE19}, and MTE~\cite{ARM2019}, \tagenc is a probabilistic mitigation technique.
A memory safety violation, such as a linear or non-linear buffer overflow, cannot be detected in \tauth by the \tagenc architecture if the color of the target memory object matches the color of the exploited memory object.
However, since \tagenc already detects a memory safety violation at the first mismatch and the attacker cannot influence the color assignment of a memory object, the attacker requires a color collision at the first try.
The probability of having a color collision of two memory objects directly corresponds to the number of used color bits.
A memory safety vulnerability, such as Heartbleed, can be detected with \tauth at the first violation with a probability of 93.7\,\% for a tag size of \SI{4}{\bits}, with a probability of 99.998\,\% for a tag size of \SI{16}{\bits}, and for a tag size of \SI{25}{\bits} even with a higher probability.
Additionally, since attacks like ROP or JOP require the adversary to build an attack chain, multiple color collisions are required increasing the detection probability.
Although larger color sizes also increase the security guarantees, schemes like ARM MTE do not utilize the full available space in the free upper bits of the pointer because storing the color in memory is required, resulting in significant memory overheads.
\tagenc prevents this security-overhead trade-off by completely avoiding storing the color in the memory, allowing the scheme to fully utilize the unused bits in the pointer and maximize security guarantees.

\paragraph{Tag Granularity.}
Due to its nature, memory coloring is an imprecise protection mechanism.
For example, when allocating a \SI{30}{\byte} memory object in \tagenc with a tag granularity of 16 byte, the full 32 byte are colored with the same color.
When accessing byte \SI{31} using a linear buffer overflow, the memory safety violation cannot be detected by any memory coloring scheme.
However, in practice, this issue can be circumvented by choosing an appropriate tag granularity.
On 64-bit \riscv systems, objects on the heap are 16 byte aligned.
Here, by choosing a tag granularity of also 16 byte, the adjacent target memory object is tagged with a different color and cannot be reached by the attacker in \tauth.
Objects on the stack are also aligned to $TG$ and the size is increased to a multiple of $TG$.
Now, the victim and target buffer, \eg a return address, are in different color domains allowing \tagenc in \tauth to detect a memory safety violation.
When the memory object size would not have been resized to a multiple of $TG$ and the tag granularity would be larger than the memory alignment, \eg $TG=64$, the same color is assigned to two, \eg \SI{32}{\byte}, adjacent memory objects making it impossible to detect an overflow.
Although a smaller $TG$ allows a more fine granular detection mechanism, it also increases the overhead for storing the colors in the cache architecture.
Similar to other research~\cite{DBLP:journals/usenix-login/Serebryany19}, we suggest to use a tag granularity of 16 byte on our reference platform.

\paragraph{Color Checking.}
Schemes like PARTS, which uses ARM's pointer authentication feature~\cite{arm2017pointerauthentication}, or CCFI~\cite{DBLP:conf/ccs/MashtizadehBBM15} use dedicated authentication instructions to verify the integrity of the pointer.
Since verification and usage is, except for dedicated instructions like the \texttt{blraa} instruction in ARM, not atomic, these countermeasures are vulnerable against time-of-check to time-of-use (TOCTOU) attacks.
\tagenc in \tauth circumvents this problem by enforcing a color check automatically in hardware for each memory access.

\paragraph{Color Management.}
Coloring a memory object with a color either can be done using a randomized or a deterministic color assignment strategy.
When using a deterministic coloring scheme, a color management mechanism needs to track the color assignment to assure that two adjacent memory objects have a different color.
An example of a system deterministically assigning tags is ARM's pointer authentication scheme, which is integrated into Apple smartphones.
However, past research~\cite{Apple2019} showed that an attacker can forge arbitrary signed pointers by using signing gadgets.
To prevent color management security issues and avoid additional overhead introduced by the mechanism, \tagenc uses a randomized coloring approach.

\section{Related Work}

This section summarizes different memory vulnerability schemes and analyzes their performance overhead and security guarantees.

\begin{table}[b]
%\vspace*{-0.45cm}
\caption{Security comparison of different memory vulnerability mitigation schemes.}
%\vspace*{-0.45cm}
\label{tab:comparison}
\begin{threeparttable}
\resizebox{\linewidth}{!}{\begin{tabular}{llllll}
\toprule
Scheme          & \begin{tabular}[c]{@{}l@{}}Code-\\ Pointer\\ Integrity\end{tabular} & \begin{tabular}[c]{@{}l@{}}Data-\\ Pointer\\ Integrity\end{tabular} & \begin{tabular}[c]{@{}l@{}}Temporal\\ Safety\end{tabular} & \begin{tabular}[c]{@{}l@{}}Spatial\\ Safety\end{tabular} & Overhead \\ 
\midrule
CCFI            & \ding{52}  & \ding{54}  & \ding{54} & \ding{54} & 52\,\%        \\
CPI             & \ding{52}  & \ding{52}  & \ding{54} & \ding{54} & 8.4\,\%         \\
PARTS           & \ding{52}  & \ding{52}  & \ding{54} & \ding{54} & 19.5\,\%         \\
SoftBound+CETS  & \ding{115} & \ding{115} & \ding{52} & \ding{52} & 116\,\%         \\
MemTagSanitizer & \faShield  & \faShield  & \ding{54} & \ding{52} & -         \\
\multirow{2}{*}{\tagenc} & \multirow{2}{*}{\ding{115}} & \multirow{2}{*}{\ding{115}} & \multirow{2}{*}{\faShield} & \multirow{2}{*}{\ding{52}} & 6.1\,\%* \\
                &        &               &              &         & $109.8\,\%^{\dag}$ \\
\bottomrule
\end{tabular}}
\small{
\begin{center}
    \ding{52} \quad Full \quad
    \faShield \quad Partial \quad
    \ding{115} \quad Indirect \quad
    \ding{54} \quad No Protection \\
    * Including Memory Encryption Overhead \\
    \dag Additional to the Memory Encryption Overhead
\end{center}
}
\end{threeparttable}
\end{table}

\subsection{Overhead Comparison}

As shown in Table~\ref{tab:comparison}, the performance overhead of less than 6.1\,\% for \tagenc on a system already featuring a memory encryption unit is low.
These numbers show that extending a system with an already integrated memory encryption scheme with \tagenc is reasonable, as memory safety can be implemented relatively cheaply.
We argue that with the increasing amount of systems providing memory encryption, such as Intel SGX or AMD TSME, also the number of platforms potentially supporting \tagenc increases.
While there is already a trend of using memory encryption in cloud computing, the announcement of Intel~\cite{IntelME} introducing memory encryption for commodity processors soon highlights the importance of it for wider deployment.
Typically, highly optimized memory encryption units can be implemented with an overhead between 5\,\% and 26\,\%~\cite{ARMMENC}.
As we do not have access to these commercial MEEs for our prototype, we used the open-source \memsec framework, where we measured an average performance overhead between 58.9\,\% and 109.8\,\%.
For the overall performance overhead of the combined physical and logical memory safety protection, we measured an overhead between 62.0\,\% and 116.1\,\%.
These numbers show that the dominating performance factor is \memsec and \emph{not} \tagenc.
Furthermore, we contend that a na\"ive combination of logical and physical memory safety, such as combining PARTS with memory encryption, accumulates both overheads.

\subsection{Security Comparison}

In general, logical memory safety strategies can be categorized into schemes that are limiting the attacker's capabilities when exploiting a memory bug (\eg PARTS, CPI, CCFI) and those detecting the exploitation of a memory safety bug.
\tagenc with \tauth uses the latter approach to thwart logical memory safety attacks by detecting a broad range of spatial and temporal memory bugs.

\paragraph{Control-flow integrity.}
Control-flow integrity~(CFI) minimizes the attacker's capability when exploiting a memory bug by limiting the control-flow of a program to only valid paths through the control-flow graph (CFG)~\cite{DBLP:conf/uss/CarliniBPWG15}.
The security of CFI schemes depends on the precision of the CFG, which is typically determined using static analysis, and the reliability of the security enforcement.
Cryptographic CFI (CCFI)~\cite{DBLP:conf/ccs/MashtizadehBBM15} improves the precision of commodity CFI schemes by dynamically performing pointer classification at runtime.
Similar to \tagenc, CCFI utilizes cryptography to enforce runtime security.
Each object that influences the control-flow of a program is tagged with the MAC over the pointer and its dynamically determined class.
The MAC is then checked before using the object.
Since computing and verifying a MAC is costly, CCFI increases the overhead by $52\,\%$.
Due to the nature of CFI schemes, CCFI cannot provide spatial and temporal memory safety.

\paragraph{Code-pointer integrity.}
Similar to CFI, code-pointer integrity (CPI)~\cite{DBLP:books/mc/18/KuznetsovSPCSS18} claims to prevent all control-flow hijack attacks, while simultaneously decreasing the performance overhead.
CPI protects sensitive code-pointers by storing them and metadata in a safe region.
While the overhead introduced by CPI is negligible, its security completely relies on the isolation of this region.
On systems without segmentation protection support, like for x86-64 systems, CPI uses information hiding to protect its safe region making it vulnerable to attacks leaking this location~\cite{DBLP:conf/sp/EvansFGOTSSRO15}.

\paragraph{Code- and data-pointer Integrity.}
Advanced attack scenarios, like ROP or DOP, show that providing data- or control-flow integrity exclusively is not sufficient.
It requires a combination of defense strategies to mitigate against a powerful attacker.
One promising attempt utilizing hardware features offered by the underlying architecture is PARTS~\cite{DBLP:conf/uss/LiljestrandNWPE19}.
PARTS implements a compiler instrumentation, which automatically adds pointer integrity checks to protect all code- and data-pointers.
Here, dedicated pointer authentication instructions are used to perform pointer signing and verification.
PARTS protects all backward-edge and forward-edge code-pointers, as well as all data-pointers.
However, the data-pointer integrity scheme does not provide temporal or spatial memory safety.
Therefore, PARTS is vulnerable against attacks targeting the data plane, like Heartbleed~\cite{DBLP:conf/imc/DurumericKAHBLWABPP14}, or other security-critical attacks on non-control data~\cite{DBLP:conf/uss/Chen0S05}.

\paragraph{Memory Safety.}
Memory safety prevents the exploitation of memory bugs.
As such, it is considered to be a stronger concept than mitigating the effects of an exploited memory bug~\cite{DBLP:conf/sp/SzekeresPWS13}.
However, software-based solutions, like the combination of SoftBound and CETS, typically yield a high performance penalty, making these schemes unrealistic to deploy on a larger scale.
To reduce the overhead, hardware support is required.
One promising hardware-assisted attempt to detect most spatial and temporal bugs is based on ARM's MTE feature is Google's MemTagSanitizer.
However, at the time of writing, MemTagSanitizer is still under development and no performance numbers are released yet and no protection of the heap is implemented.
Although we expect that this scheme will provide similar performance than \tagenc, MTE only provides restricted security guarantees.
MTE uses a small tag size to limit the memory overhead introduced by storing the tags in memory.

\section{Conclusion \& Future Work}
\label{sec:conclusion}
Current memory security schemes either are incomplete~\cite{DBLP:conf/uss/LiljestrandNWPE19,DBLP:conf/ccs/MashtizadehBBM15}, do not provide enough security~\cite{ARM2019,DBLP:books/mc/18/KuznetsovSPCSS18}, or add non-negligible overhead~\cite{DBLP:conf/pldi/NagarakatteZMZ09,DBLP:conf/iwmm/NagarakatteZMZ10}, especially when combined with physical memory safety, to the system.
\tagenc closes these gaps by introducing a memory safety concept based on a hardware-assisted memory coloring scheme.
\tagenc combines memory encryption with memory coloring to thwart a broad range of physical and logical memory safety vulnerabilities.
By combining these two mechanisms, we show that memory coloring almost comes for free and memory safety vulnerabilities can efficiently be protected.
The design uses a color, stored inside the related pointer, and propagates this value up to the data cache of the system.
This color value is used to tweak the memory encryption system, thus avoiding storing the color in memory.
Our approach shows that the performance overhead for \tagenc is negligible and, therefore, can be used for large scale deployment.
In this paper, we provide an end-to-end solution from the concept to the prototype implementation of our design.
We integrated \tagenc to a \riscv based processing platform and adapted an LLVM toolchain and developed a runtime library to automatically instrument programs and protect all memory allocations of the application without the need for user annotations.
Our evaluation shows that the hardware overhead for these changes is less than 1\,\% and the software overhead compared to a system already featuring a memory encryption unit is on average less than 6.1\,\%, which makes this design practical for real-life applications.

\paragraph{Future Work.}
As mentioned in Section~\ref{sec:plim}, the performance of \tagenc largely depends on the performance of the memory encryption engine~(MEE).
Hence, a possible future work would be to optimize the performance of \memsec.
Currently, \memsec operates at the same clock frequency as the processor core.
To increase the memory bandwidth and further decrease the latency of memory accesses, \memsec could be placed next to the memory to operate on a much higher clock frequency.
However, this requires to optimize the inner logic of \memsec to avoid any timing violations.
Currently, \memsec is a highly flexible framework allowing several corner cases, such as AXI bursts and strobes.
Here, one strategy to maximize the performance of the MEE could be to adapt \memsec to the target architecture and remove functionalities not supported by this architecture.
If providing physical memory safety is not needed, a final optimization step could be to only encrypt colored memory objects and bypass the MEE for non-colored objects.

\begin{acks}
We thank Samuel Weiser, David Schrammel, and the anonymous reviewers for helping to improve this paper.
This project has received funding from the European Research Council (ERC) under the European Union’s Horizon 2020 research and innovation programme (grant agreement No 681402) and by the Austrian Research Promotion Agency (FFG) via the competence center Know-Center (grant number 844595), which is funded in the context of COMET - Competence Centers for Excellent Technologies by BMVIT, BMWFW, and Styria.
\end{acks}

\bibliographystyle{ACM-Reference-Format}
\balance
\bibliography{main}


\begin{thebibliography}{64}

%%% the Web address.

\ifx \showCODEN    \undefined \def \showCODEN     #1{\unskip}     \fi
\ifx \showDOI      \undefined \def \showDOI       #1{#1}\fi
\ifx \showISBNx    \undefined \def \showISBNx     #1{\unskip}     \fi
\ifx \showISBNxiii \undefined \def \showISBNxiii  #1{\unskip}     \fi
\ifx \showISSN     \undefined \def \showISSN      #1{\unskip}     \fi
\ifx \showLCCN     \undefined \def \showLCCN      #1{\unskip}     \fi
\ifx \shownote     \undefined \def \shownote      #1{#1}          \fi
\ifx \showarticletitle \undefined \def \showarticletitle #1{#1}   \fi
\ifx \showURL      \undefined \def \showURL       {\relax}        \fi
% invisible to TeX
\providecommand\bibfield[2]{#2}
\providecommand\bibinfo[2]{#2}
\providecommand\natexlab[1]{#1}
\providecommand\showeprint[2][]{arXiv:#2}

\bibitem[\protect\citeauthoryear{Aingaran, Jairath, Konstadinidis, Leung,
  Loewenstein, McAllister, Phillips, Radovic, Sivaramakrishnan, Smentek, and
  Wicki}{Aingaran et~al\mbox{.}}{2015}]%
        {DBLP:journals/micro/AingaranJKLLMPR15}
\bibfield{author}{\bibinfo{person}{Kathirgamar Aingaran},
  \bibinfo{person}{Sumti Jairath}, \bibinfo{person}{Georgios~K. Konstadinidis},
  \bibinfo{person}{Serena Leung}, \bibinfo{person}{Paul Loewenstein},
  \bibinfo{person}{Curtis McAllister}, \bibinfo{person}{Stephen Phillips},
  \bibinfo{person}{Zoran Radovic}, \bibinfo{person}{Ram Sivaramakrishnan},
  \bibinfo{person}{David Smentek}, {and} \bibinfo{person}{Thomas Wicki}.}
  \bibinfo{year}{2015}\natexlab{}.
\newblock \showarticletitle{\href{https://doi.org/10.1109/MM.2015.35}{{M7:}
  Oracle's Next-Generation Sparc Processor}}.
\newblock \bibinfo{journal}{\emph{{IEEE} Micro}}  \bibinfo{volume}{35}
  (\bibinfo{year}{2015}).
\newblock


\bibitem[\protect\citeauthoryear{ARM}{ARM}{2019}]%
        {ARM2019}
\bibfield{author}{\bibinfo{person}{ARM}.} \bibinfo{year}{2019}\natexlab{}.
\newblock
  \bibinfo{booktitle}{\emph{\href{https://developer.arm.com/-/media/Arm\%20Developer\%20Community/PDF/Arm_Memory_Tagging_Extension_Whitepaper.pdf}{Armv8.5-A
  Memory Tagging Extension}}}.
\newblock


\bibitem[\protect\citeauthoryear{ARM}{ARM}{2020}]%
        {ARM82020}
\bibfield{author}{\bibinfo{person}{ARM}.} \bibinfo{year}{2020}\natexlab{}.
\newblock
  \bibinfo{booktitle}{\emph{\href{https://static.docs.arm.com/ddi0487/fb/DDI0487F_b_armv8_arm.pdf}{Arm
  Architecture Reference Manual}}}.
\newblock


\bibitem[\protect\citeauthoryear{Avanzi}{Avanzi}{2016}]%
        {DBLP:journals/iacr/Avanzi16}
\bibfield{author}{\bibinfo{person}{Roberto Avanzi}.}
  \bibinfo{year}{2016}\natexlab{}.
\newblock \showarticletitle{\href{http://eprint.iacr.org/2016/444}{The {QARMA}
  Block Cipher Family - Almost {MDS} Matrices Over Rings With Zero Divisors,
  Nearly Symmetric Even-Mansour Constructions With Non-Involutory Central
  Rounds, and Search Heuristics for Low-Latency S-Boxes}}.
\newblock \bibinfo{journal}{\emph{ePrint 2016/444}} (\bibinfo{year}{2016}).
\newblock


\bibitem[\protect\citeauthoryear{Azad}{Azad}{2019}]%
        {Apple2019}
\bibfield{author}{\bibinfo{person}{Brandon Azad}.}
  \bibinfo{year}{2019}\natexlab{}.
\newblock
  \bibinfo{booktitle}{\emph{\href{https://googleprojectzero.blogspot.com/2019/02/examining-pointer-authentication-on.html}{Examining
  Pointer Authentication on the iPhone XS}}}.
\newblock


\bibitem[\protect\citeauthoryear{Carlini, Barresi, Payer, Wagner, and
  Gross}{Carlini et~al\mbox{.}}{2015}]%
        {DBLP:conf/uss/CarliniBPWG15}
\bibfield{author}{\bibinfo{person}{Nicholas Carlini}, \bibinfo{person}{Antonio
  Barresi}, \bibinfo{person}{Mathias Payer}, \bibinfo{person}{David~A. Wagner},
  {and} \bibinfo{person}{Thomas~R. Gross}.} \bibinfo{year}{2015}\natexlab{}.
\newblock
  \showarticletitle{\href{https://www.usenix.org/conference/usenixsecurity15/technical-sessions/presentation/carlini}{Control-Flow
  Bending: On the Effectiveness of Control-Flow Integrity}}. In
  \bibinfo{booktitle}{\emph{{USENIX} Security Symposium}}.
\newblock


\bibitem[\protect\citeauthoryear{Chen, Xu, and Sezer}{Chen
  et~al\mbox{.}}{2005}]%
        {DBLP:conf/uss/Chen0S05}
\bibfield{author}{\bibinfo{person}{Shuo Chen}, \bibinfo{person}{Jun Xu}, {and}
  \bibinfo{person}{Emre~Can Sezer}.} \bibinfo{year}{2005}\natexlab{}.
\newblock
  \showarticletitle{\href{https://www.usenix.org/conference/14th-usenix-security-symposium/non-control-data-attacks-are-realistic-threats}{Non-Control-Data
  Attacks Are Realistic Threats}}. In \bibinfo{booktitle}{\emph{{USENIX}
  Security Symposium}}.
\newblock


\bibitem[\protect\citeauthoryear{Clark and Corrigan}{Clark and
  Corrigan}{1989}]%
        {DBLP:journals/ibmsj/ClarkC89}
\bibfield{author}{\bibinfo{person}{Brian~E. Clark} {and}
  \bibinfo{person}{Michael~J. Corrigan}.} \bibinfo{year}{1989}\natexlab{}.
\newblock
  \showarticletitle{\href{https://doi.org/10.1147/sj.283.0407}{Application
  System/400 Performance Characteristics}}.
\newblock \bibinfo{journal}{\emph{{IBM} Syst. J.}}  \bibinfo{volume}{28}
  (\bibinfo{year}{1989}).
\newblock


\bibitem[\protect\citeauthoryear{Corporation}{Corporation}{2019}]%
        {Corporation2019}
\bibfield{author}{\bibinfo{person}{Intel Corporation}.}
  \bibinfo{year}{2019}\natexlab{}.
\newblock
  \bibinfo{booktitle}{\emph{\href{https://software.intel.com/sites/default/files/managed/a5/16/Multi-Key-Total-Memory-Encryption-Spec.pdf}{{Intel{\textregistered}
  Architecture Memory Encryption Technologies Specification}}}}.
\newblock \bibinfo{type}{{T}echnical {R}eport}.
\newblock


\bibitem[\protect\citeauthoryear{Dhawan, Vasilakis, Rubin, Chiricescu, Smith,
  Jr., Pierce, and DeHon}{Dhawan et~al\mbox{.}}{2014}]%
        {DBLP:conf/isca/DhawanVRCSKPD14}
\bibfield{author}{\bibinfo{person}{Udit Dhawan}, \bibinfo{person}{Nikos
  Vasilakis}, \bibinfo{person}{Raphael Rubin}, \bibinfo{person}{Silviu
  Chiricescu}, \bibinfo{person}{Jonathan~M. Smith}, \bibinfo{person}{Thomas
  F.~Knight Jr.}, \bibinfo{person}{Benjamin~C. Pierce}, {and}
  \bibinfo{person}{Andr{\'{e}} DeHon}.} \bibinfo{year}{2014}\natexlab{}.
\newblock
  \showarticletitle{\href{https://doi.org/10.1145/2611765.2611773}{{PUMP:} a
  programmable unit for metadata processing}}. In
  \bibinfo{booktitle}{\emph{International Symposium on Computer Architecture --
  {ISCA}}}.
\newblock


\bibitem[\protect\citeauthoryear{Dobraunig, Eichlseder, Mendel, and
  Schl{\"{a}}ffer}{Dobraunig et~al\mbox{.}}{2016}]%
        {Dobraunig2016}
\bibfield{author}{\bibinfo{person}{Christoph Dobraunig}, \bibinfo{person}{Maria
  Eichlseder}, \bibinfo{person}{Florian Mendel}, {and} \bibinfo{person}{Martin
  Schl{\"{a}}ffer}.} \bibinfo{year}{2016}\natexlab{}.
\newblock \bibinfo{booktitle}{\emph{\href{http://ascon.iaik.tugraz.at}{{Ascon
  v1.2 Submission to the CAESAR Competition}}}}.
\newblock \bibinfo{type}{{T}echnical {R}eport}.
\newblock


\bibitem[\protect\citeauthoryear{Duck and Yap}{Duck and Yap}{2016}]%
        {DBLP:conf/cc/DuckY16}
\bibfield{author}{\bibinfo{person}{Gregory~J. Duck} {and}
  \bibinfo{person}{Roland H.~C. Yap}.} \bibinfo{year}{2016}\natexlab{}.
\newblock \showarticletitle{\href{https://doi.org/10.1145/2892208.2892212}{Heap
  bounds protection with low fat pointers}}. In
  \bibinfo{booktitle}{\emph{Proceedings of the 25th International Conference on
  Compiler Construction, {CC} 2016, Barcelona, Spain, March 12-18, 2016}}.
\newblock


\bibitem[\protect\citeauthoryear{Durumeric, Kasten, Adrian, Halderman, Bailey,
  Li, Weaver, Amann, Beekman, Payer, and Paxson}{Durumeric
  et~al\mbox{.}}{2014}]%
        {DBLP:conf/imc/DurumericKAHBLWABPP14}
\bibfield{author}{\bibinfo{person}{Zakir Durumeric}, \bibinfo{person}{James
  Kasten}, \bibinfo{person}{David Adrian}, \bibinfo{person}{J.~Alex Halderman},
  \bibinfo{person}{Michael Bailey}, \bibinfo{person}{Frank Li},
  \bibinfo{person}{Nicholas Weaver}, \bibinfo{person}{Johanna Amann},
  \bibinfo{person}{Jethro Beekman}, \bibinfo{person}{Mathias Payer}, {and}
  \bibinfo{person}{Vern Paxson}.} \bibinfo{year}{2014}\natexlab{}.
\newblock \showarticletitle{\href{https://doi.org/10.1145/2663716.2663755}{The
  Matter of Heartbleed}}. In \bibinfo{booktitle}{\emph{Internet Measurement
  Conference -- {IMC}}}.
\newblock


\bibitem[\protect\citeauthoryear{Elbaz, Champagne, Gebotys, Lee, Potlapally,
  and Torres}{Elbaz et~al\mbox{.}}{2009}]%
        {DBLP:journals/tcos/ElbazCGLPT09}
\bibfield{author}{\bibinfo{person}{Reouven Elbaz}, \bibinfo{person}{David
  Champagne}, \bibinfo{person}{Catherine~H. Gebotys}, \bibinfo{person}{Ruby~B.
  Lee}, \bibinfo{person}{Nachiketh~R. Potlapally}, {and}
  \bibinfo{person}{Lionel Torres}.} \bibinfo{year}{2009}\natexlab{}.
\newblock
  \showarticletitle{\href{https://doi.org/10.1007/978-3-642-01004-0_1}{Hardware
  Mechanisms for Memory Authentication: {A} Survey of Existing Techniques and
  Engines}}.
\newblock \bibinfo{journal}{\emph{Trans. Comput. Sci.}}  \bibinfo{volume}{4}
  (\bibinfo{year}{2009}).
\newblock


\bibitem[\protect\citeauthoryear{Evans, Fingeret, Gonzalez, Otgonbaatar, Tang,
  Shrobe, Sidiroglou{-}Douskos, Rinard, and Okhravi}{Evans
  et~al\mbox{.}}{2015}]%
        {DBLP:conf/sp/EvansFGOTSSRO15}
\bibfield{author}{\bibinfo{person}{Isaac Evans}, \bibinfo{person}{Sam
  Fingeret}, \bibinfo{person}{Julian Gonzalez}, \bibinfo{person}{Ulziibayar
  Otgonbaatar}, \bibinfo{person}{Tiffany Tang}, \bibinfo{person}{Howard~E.
  Shrobe}, \bibinfo{person}{Stelios Sidiroglou{-}Douskos},
  \bibinfo{person}{Martin Rinard}, {and} \bibinfo{person}{Hamed Okhravi}.}
  \bibinfo{year}{2015}\natexlab{}.
\newblock \showarticletitle{\href{https://doi.org/10.1109/SP.2015.53}{Missing
  the Point(er): On the Effectiveness of Code Pointer Integrity}}. In
  \bibinfo{booktitle}{\emph{{IEEE} Symposium on Security and Privacy --
  {S{\&}P}}}.
\newblock


\bibitem[\protect\citeauthoryear{Feustel}{Feustel}{1972}]%
        {DBLP:conf/afips/Feustel72}
\bibfield{author}{\bibinfo{person}{Edward~A. Feustel}.}
  \bibinfo{year}{1972}\natexlab{}.
\newblock \showarticletitle{\href{https://doi.org/10.1145/1478873.1478920}{The
  Rice research computer: a tagged architecture}}. In
  \bibinfo{booktitle}{\emph{American Federation of Information Processing
  Societies -- {AFIPS}}}.
\newblock


\bibitem[\protect\citeauthoryear{Gueron}{Gueron}{2016}]%
        {DBLP:journals/iacr/Gueron16}
\bibfield{author}{\bibinfo{person}{Shay Gueron}.}
  \bibinfo{year}{2016}\natexlab{}.
\newblock \showarticletitle{\href{http://eprint.iacr.org/2016/204}{A Memory
  Encryption Engine Suitable for General Purpose Processors}}.
\newblock \bibinfo{journal}{\emph{ePrint 2016/204}} (\bibinfo{year}{2016}).
\newblock


\bibitem[\protect\citeauthoryear{Guthaus, Ringenberg, Ernst, Austin, Mudge, and
  Brown}{Guthaus et~al\mbox{.}}{2001}]%
        {guthaus2001mibench}
\bibfield{author}{\bibinfo{person}{Matthew~R Guthaus},
  \bibinfo{person}{Jeffrey~S Ringenberg}, \bibinfo{person}{Dan Ernst},
  \bibinfo{person}{Todd~M Austin}, \bibinfo{person}{Trevor Mudge}, {and}
  \bibinfo{person}{Richard~B Brown}.} \bibinfo{year}{2001}\natexlab{}.
\newblock \showarticletitle{MiBench: A free, commercially representative
  embedded benchmark suite}. In \bibinfo{booktitle}{\emph{Proceedings of the
  fourth annual IEEE international workshop on workload characterization. WWC-4
  (Cat. No. 01EX538)}}.
\newblock


\bibitem[\protect\citeauthoryear{Halderman, Schoen, Heninger, Clarkson, Paul,
  Calandrino, Feldman, Appelbaum, and Felten}{Halderman et~al\mbox{.}}{2008}]%
        {DBLP:conf/uss/HaldermanSHCPCFAF08}
\bibfield{author}{\bibinfo{person}{J.~Alex Halderman}, \bibinfo{person}{Seth~D.
  Schoen}, \bibinfo{person}{Nadia Heninger}, \bibinfo{person}{William
  Clarkson}, \bibinfo{person}{William Paul}, \bibinfo{person}{Joseph~A.
  Calandrino}, \bibinfo{person}{Ariel~J. Feldman}, \bibinfo{person}{Jacob
  Appelbaum}, {and} \bibinfo{person}{Edward~W. Felten}.}
  \bibinfo{year}{2008}\natexlab{}.
\newblock
  \showarticletitle{\href{http://www.usenix.org/events/sec08/tech/full_papers/halderman/halderman.pdf}{Lest
  We Remember: Cold Boot Attacks on Encryption Keys}}. In
  \bibinfo{booktitle}{\emph{{USENIX} Security Symposium}}.
\newblock


\bibitem[\protect\citeauthoryear{Hu, Shinde, Adrian, Chua, Saxena, and
  Liang}{Hu et~al\mbox{.}}{2016}]%
        {DBLP:conf/sp/HuSACSL16}
\bibfield{author}{\bibinfo{person}{Hong Hu}, \bibinfo{person}{Shweta Shinde},
  \bibinfo{person}{Sendroiu Adrian}, \bibinfo{person}{Zheng~Leong Chua},
  \bibinfo{person}{Prateek Saxena}, {and} \bibinfo{person}{Zhenkai Liang}.}
  \bibinfo{year}{2016}\natexlab{}.
\newblock
  \showarticletitle{\href{https://doi.org/10.1109/SP.2016.62}{Data-Oriented
  Programming: On the Expressiveness of Non-control Data Attacks}}. In
  \bibinfo{booktitle}{\emph{{IEEE} Symposium on Security and Privacy --
  {S{\&}P}}}.
\newblock


\bibitem[\protect\citeauthoryear{Joannou, Woodruff, Kovacsics, Moore, Bradbury,
  Xia, Watson, Chisnall, Roe, Davis, Napierala, Baldwin, Gudka, Neumann,
  Mazzinghi, Richardson, Son, and Markettos}{Joannou et~al\mbox{.}}{2017}]%
        {DBLP:conf/iccd/JoannouWKMBXWCR17}
\bibfield{author}{\bibinfo{person}{Alexandre Joannou},
  \bibinfo{person}{Jonathan Woodruff}, \bibinfo{person}{Robert Kovacsics},
  \bibinfo{person}{Simon~W. Moore}, \bibinfo{person}{Alex Bradbury},
  \bibinfo{person}{Hongyan Xia}, \bibinfo{person}{Robert N.~M. Watson},
  \bibinfo{person}{David Chisnall}, \bibinfo{person}{Michael Roe},
  \bibinfo{person}{Brooks Davis}, \bibinfo{person}{Edward Napierala},
  \bibinfo{person}{John Baldwin}, \bibinfo{person}{Khilan Gudka},
  \bibinfo{person}{Peter~G. Neumann}, \bibinfo{person}{Alfredo Mazzinghi},
  \bibinfo{person}{Alex Richardson}, \bibinfo{person}{Stacey~D. Son}, {and}
  \bibinfo{person}{A.~Theodore Markettos}.} \bibinfo{year}{2017}\natexlab{}.
\newblock
  \showarticletitle{\href{https://doi.org/10.1109/ICCD.2017.112}{Efficient
  Tagged Memory}}. In \bibinfo{booktitle}{\emph{International Conference on
  Computer Design -- {ICCD}}}.
\newblock


\bibitem[\protect\citeauthoryear{Kaplan, Powell, and Tom}{Kaplan
  et~al\mbox{.}}{2016}]%
        {Kaplan2016a}
\bibfield{author}{\bibinfo{person}{David Kaplan}, \bibinfo{person}{Jeremy
  Powell}, {and} \bibinfo{person}{Woller Tom}.}
  \bibinfo{year}{2016}\natexlab{}.
\newblock
  \bibinfo{title}{\href{https://developer.amd.com/wordpress/media/2013/12/AMD_Memory_Encryption_Whitepaper_v7-Public.pdf}{{AMD
  MEMORY ENCRYPTION}}}.  (\bibinfo{year}{2016}).
\newblock


\bibitem[\protect\citeauthoryear{Kim, Daly, Kim, Fallin, Lee, Lee, Wilkerson,
  Lai, and Mutlu}{Kim et~al\mbox{.}}{2014}]%
        {DBLP:conf/isca/KimDKFLLWLM14}
\bibfield{author}{\bibinfo{person}{Yoongu Kim}, \bibinfo{person}{Ross Daly},
  \bibinfo{person}{Jeremie Kim}, \bibinfo{person}{Chris Fallin},
  \bibinfo{person}{Ji{-}Hye Lee}, \bibinfo{person}{Donghyuk Lee},
  \bibinfo{person}{Chris Wilkerson}, \bibinfo{person}{Konrad Lai}, {and}
  \bibinfo{person}{Onur Mutlu}.} \bibinfo{year}{2014}\natexlab{}.
\newblock
  \showarticletitle{\href{https://doi.org/10.1109/ISCA.2014.6853210}{Flipping
  bits in memory without accessing them: An experimental study of {DRAM}
  disturbance errors}}. In \bibinfo{booktitle}{\emph{International Symposium on
  Computer Architecture -- {ISCA}}}.
\newblock


\bibitem[\protect\citeauthoryear{Kuznetsov, Szekeres, Payer, Candea, Sekar, and
  Song}{Kuznetsov et~al\mbox{.}}{2018}]%
        {DBLP:books/mc/18/KuznetsovSPCSS18}
\bibfield{author}{\bibinfo{person}{Volodymyr Kuznetsov},
  \bibinfo{person}{Laszlo Szekeres}, \bibinfo{person}{Mathias Payer},
  \bibinfo{person}{George Candea}, \bibinfo{person}{R. Sekar}, {and}
  \bibinfo{person}{Dawn Song}.} \bibinfo{year}{2018}\natexlab{}.
\newblock
  \showarticletitle{\href{https://doi.org/10.1145/3129743.3129748}{Code-pointer
  integrity}}.
\newblock In \bibinfo{booktitle}{\emph{The Continuing Arms Race: Code-Reuse
  Attacks and Defenses}}.
\newblock


\bibitem[\protect\citeauthoryear{Kwon, Dhawan, Smith, Jr., and DeHon}{Kwon
  et~al\mbox{.}}{2013}]%
        {DBLP:conf/ccs/KwonDSKD13}
\bibfield{author}{\bibinfo{person}{Albert Kwon}, \bibinfo{person}{Udit Dhawan},
  \bibinfo{person}{Jonathan~M. Smith}, \bibinfo{person}{Thomas F.~Knight Jr.},
  {and} \bibinfo{person}{Andr{\'{e}} DeHon}.} \bibinfo{year}{2013}\natexlab{}.
\newblock
  \showarticletitle{\href{https://doi.org/10.1145/2508859.2516713}{Low-fat
  pointers: compact encoding and efficient gate-level implementation of fat
  pointers for spatial safety and capability-based security}}. In
  \bibinfo{booktitle}{\emph{Conference on Computer and Communications Security
  -- {CCS}}}.
\newblock


\bibitem[\protect\citeauthoryear{Lattner and Adve}{Lattner and Adve}{2004}]%
        {DBLP:conf/cgo/LattnerA04}
\bibfield{author}{\bibinfo{person}{Chris Lattner} {and}
  \bibinfo{person}{Vikram~S. Adve}.} \bibinfo{year}{2004}\natexlab{}.
\newblock
  \showarticletitle{\href{https://doi.org/10.1109/CGO.2004.1281665}{{LLVM:} {A}
  Compilation Framework for Lifelong Program Analysis {\&} Transformation}}. In
  \bibinfo{booktitle}{\emph{International Symposium on Code Generation and
  Optimization -- {CGO}}}.
\newblock


\bibitem[\protect\citeauthoryear{Lee, Kohlbrenner, Shinde, Song, and
  Asanovic}{Lee et~al\mbox{.}}{2019}]%
        {DBLP:journals/corr/abs-1907-10119}
\bibfield{author}{\bibinfo{person}{Dayeol Lee}, \bibinfo{person}{David
  Kohlbrenner}, \bibinfo{person}{Shweta Shinde}, \bibinfo{person}{Dawn Song},
  {and} \bibinfo{person}{Krste Asanovic}.} \bibinfo{year}{2019}\natexlab{}.
\newblock \showarticletitle{\href{http://arxiv.org/abs/1907.10119}{Keystone:
  {A} Framework for Architecting TEEs}}.
\newblock \bibinfo{journal}{\emph{arXiv abs/1907.10119}}
  (\bibinfo{year}{2019}).
\newblock


\bibitem[\protect\citeauthoryear{Liljestrand, Nyman, Wang, Perez, Ekberg, and
  Asokan}{Liljestrand et~al\mbox{.}}{2019}]%
        {DBLP:conf/uss/LiljestrandNWPE19}
\bibfield{author}{\bibinfo{person}{Hans Liljestrand}, \bibinfo{person}{Thomas
  Nyman}, \bibinfo{person}{Kui Wang}, \bibinfo{person}{Carlos~Chinea Perez},
  \bibinfo{person}{Jan{-}Erik Ekberg}, {and} \bibinfo{person}{N. Asokan}.}
  \bibinfo{year}{2019}\natexlab{}.
\newblock
  \showarticletitle{\href{https://www.usenix.org/conference/usenixsecurity19/presentation/liljestrand}{{PAC}
  it up: Towards Pointer Integrity using {ARM} Pointer Authentication}}. In
  \bibinfo{booktitle}{\emph{{USENIX} Security Symposium}}.
\newblock


\bibitem[\protect\citeauthoryear{Marinas}{Marinas}{2020}]%
        {LinuxArm20}
\bibfield{author}{\bibinfo{person}{Catalin Marinas}.}
  \bibinfo{year}{2020}\natexlab{}.
\newblock
  \bibinfo{booktitle}{\emph{\href{https://www.kernel.org/doc/html/latest/arm64/memory.html}{Memory
  Layout on AArch64 Linux}}}.
\newblock


\bibitem[\protect\citeauthoryear{Mashtizadeh, Bittau, Boneh, and
  Mazi{\`{e}}res}{Mashtizadeh et~al\mbox{.}}{2015}]%
        {DBLP:conf/ccs/MashtizadehBBM15}
\bibfield{author}{\bibinfo{person}{Ali~Jos{\'{e}} Mashtizadeh},
  \bibinfo{person}{Andrea Bittau}, \bibinfo{person}{Dan Boneh}, {and}
  \bibinfo{person}{David Mazi{\`{e}}res}.} \bibinfo{year}{2015}\natexlab{}.
\newblock
  \showarticletitle{\href{https://doi.org/10.1145/2810103.2813676}{{CCFI:}
  Cryptographically Enforced Control Flow Integrity}}. In
  \bibinfo{booktitle}{\emph{Conference on Computer and Communications Security
  -- {CCS}}}.
\newblock


\bibitem[\protect\citeauthoryear{Mayer}{Mayer}{1982}]%
        {Mayer1982}
\bibfield{author}{\bibinfo{person}{Alastair J.~W. Mayer}.}
  \bibinfo{year}{1982}\natexlab{}.
\newblock \showarticletitle{\href{https://doi.org/10.1145/641542.641543}{The
  Architecture of the Burroughs B5000: 20 Years Later and Still Ahead of the
  Times?}}
\newblock \bibinfo{journal}{\emph{SIGARCH Comput. Archit. News}}
  \bibinfo{volume}{10} (\bibinfo{year}{1982}).
\newblock


\bibitem[\protect\citeauthoryear{McKeen, Alexandrovich, Anati, Caspi, Johnson,
  Leslie-Hurd, and Rozas}{McKeen et~al\mbox{.}}{2016}]%
        {2016sgx}
\bibfield{author}{\bibinfo{person}{Frank McKeen}, \bibinfo{person}{Ilya
  Alexandrovich}, \bibinfo{person}{Ittai Anati}, \bibinfo{person}{Dror Caspi},
  \bibinfo{person}{Simon Johnson}, \bibinfo{person}{Rebekah Leslie-Hurd}, {and}
  \bibinfo{person}{Carlos Rozas}.} \bibinfo{year}{2016}\natexlab{}.
\newblock
  \showarticletitle{\href{https://doi.org/10.1145/2948618.2954331}{Intel®
  Software Guard Extensions (Intel® SGX) Support for Dynamic Memory Management
  Inside an Enclave}}. In \bibinfo{booktitle}{\emph{Proceedings of the Hardware
  and Architectural Support for Security and Privacy 2016}}.
  \bibinfo{publisher}{Association for Computing Machinery}.
\newblock


\bibitem[\protect\citeauthoryear{McVoy and Staelin}{McVoy and Staelin}{1996}]%
        {DBLP:conf/usenix/McVoyS96}
\bibfield{author}{\bibinfo{person}{Larry~W. McVoy} {and} \bibinfo{person}{Carl
  Staelin}.} \bibinfo{year}{1996}\natexlab{}.
\newblock \showarticletitle{lmbench: Portable Tools for Performance Analysis}.
  In \bibinfo{booktitle}{\emph{{USENIX} Annual Technical Conference}}.
\newblock


\bibitem[\protect\citeauthoryear{Miller}{Miller}{2019}]%
        {Microsoft19}
\bibfield{author}{\bibinfo{person}{Matt Miller}.}
  \bibinfo{year}{2019}\natexlab{}.
\newblock \showarticletitle{{Trends, Challanges, and Strategic Shifts in the
  Software Vulnerability Mitigation Landscape}}.
\newblock \bibinfo{journal}{\emph{BlueHat IL}} (\bibinfo{year}{2019}).
\newblock


\bibitem[\protect\citeauthoryear{MITRE}{MITRE}{2019}]%
        {Mitre2019}
\bibfield{author}{\bibinfo{person}{MITRE}.} \bibinfo{year}{2019}\natexlab{}.
\newblock
  \bibinfo{booktitle}{\emph{\href{http://cwe.mitre.org/top25/archive/2019/2019_cwe_top25.html}{CWE
  Top 25 Most Dangerous Software Errors}}}.
\newblock


\bibitem[\protect\citeauthoryear{Nagarakatte, Martin, and
  Zdancewic}{Nagarakatte et~al\mbox{.}}{2012}]%
        {DBLP:conf/isca/NagarakatteMZ12}
\bibfield{author}{\bibinfo{person}{Santosh Nagarakatte}, \bibinfo{person}{Milo
  M.~K. Martin}, {and} \bibinfo{person}{Steve Zdancewic}.}
  \bibinfo{year}{2012}\natexlab{}.
\newblock
  \showarticletitle{\href{https://doi.org/10.1109/ISCA.2012.6237017}{Watchdog:
  Hardware for safe and secure manual memory management and full memory
  safety}}. In \bibinfo{booktitle}{\emph{International Symposium on Computer
  Architecture -- {ISCA}}}.
\newblock


\bibitem[\protect\citeauthoryear{Nagarakatte, Zhao, Martin, and
  Zdancewic}{Nagarakatte et~al\mbox{.}}{2009}]%
        {DBLP:conf/pldi/NagarakatteZMZ09}
\bibfield{author}{\bibinfo{person}{Santosh Nagarakatte},
  \bibinfo{person}{Jianzhou Zhao}, \bibinfo{person}{Milo M.~K. Martin}, {and}
  \bibinfo{person}{Steve Zdancewic}.} \bibinfo{year}{2009}\natexlab{}.
\newblock
  \showarticletitle{\href{https://doi.org/10.1145/1542476.1542504}{SoftBound:
  highly compatible and complete spatial memory safety for c}}. In
  \bibinfo{booktitle}{\emph{Programming Language Design and Implementation --
  {PLDI}}}.
\newblock


\bibitem[\protect\citeauthoryear{Nagarakatte, Zhao, Martin, and
  Zdancewic}{Nagarakatte et~al\mbox{.}}{2010}]%
        {DBLP:conf/iwmm/NagarakatteZMZ10}
\bibfield{author}{\bibinfo{person}{Santosh Nagarakatte},
  \bibinfo{person}{Jianzhou Zhao}, \bibinfo{person}{Milo M.~K. Martin}, {and}
  \bibinfo{person}{Steve Zdancewic}.} \bibinfo{year}{2010}\natexlab{}.
\newblock
  \showarticletitle{\href{https://doi.org/10.1145/1806651.1806657}{{CETS:}
  compiler enforced temporal safety for {C}}}. In
  \bibinfo{booktitle}{\emph{International Symposium on Memory Management --
  {ISMM}}}.
\newblock


\bibitem[\protect\citeauthoryear{Pozo and Miller}{Pozo and Miller}{[n.d.]}]%
        {pozo98scimark}
\bibfield{author}{\bibinfo{person}{Roldan Pozo} {and} \bibinfo{person}{Bruce
  Miller}.} \bibinfo{year}{[n.d.]}\natexlab{}.
\newblock
  \bibinfo{booktitle}{\emph{\href{http://math.nist.gov/scimark2}{Scimark 2}}}.
\newblock


\bibitem[\protect\citeauthoryear{Project}{Project}{2020}]%
        {MemTagSan2020}
\bibfield{author}{\bibinfo{person}{LLVM Project}.}
  \bibinfo{year}{2020}\natexlab{}.
\newblock
  \bibinfo{booktitle}{\emph{\href{https://llvm.org/docs/MemTagSanitizer.html}{MemTagSanitizer}}}.
\newblock


\bibitem[\protect\citeauthoryear{Qualcomm~Technologies}{Qualcomm~Technologies}{2017}]%
        {arm2017pointerauthentication}
\bibfield{author}{\bibinfo{person}{Inc. Qualcomm~Technologies}.}
  \bibinfo{year}{2017}\natexlab{}.
\newblock
  \bibinfo{booktitle}{\emph{\href{https://www.qualcomm.com/media/documents/files/whitepaper-pointer-authentication-on-armv8-3.pdf}{Pointer
  Authentication on ARMv8.3}}}.
\newblock


\bibitem[\protect\citeauthoryear{Roberto-Maria}{Roberto-Maria}{2020}]%
        {ARMMENC}
\bibfield{author}{\bibinfo{person}{Avanzi Roberto-Maria}.}
  \bibinfo{year}{2020}\natexlab{}.
\newblock
  \bibinfo{booktitle}{\emph{\href{https://rwc.iacr.org/2020/slides/Avanzi.pdf}{Memory
  Protection for the ARM Architecture}}}.
\newblock


\bibitem[\protect\citeauthoryear{SALTER}{SALTER}{2020}]%
        {ars}
\bibfield{author}{\bibinfo{person}{JIM SALTER}.}
  \bibinfo{year}{2020}\natexlab{}.
\newblock
  \bibinfo{booktitle}{\emph{\href{https://arstechnica.com/gadgets/2020/02/intel-promises-full-memory-encryption-in-upcoming-cpus}{Intel
  promises Full Memory Encryption in upcoming CPUs}}}.
\newblock


\bibitem[\protect\citeauthoryear{Salter}{Salter}{2020}]%
        {IntelME}
\bibfield{author}{\bibinfo{person}{Jim Salter}.}
  \bibinfo{year}{2020}\natexlab{}.
\newblock
  \bibinfo{booktitle}{\emph{\href{https://arstechnica.com/gadgets/2020/02/intel-promises-full-memory-encryption-in-upcoming-cpus/}{Intel
  promises Full Memory Encryption in upcoming CPUs}}}.
\newblock


\bibitem[\protect\citeauthoryear{Schilling, Werner, Nasahl, and
  Mangard}{Schilling et~al\mbox{.}}{2018}]%
        {DBLP:conf/acsac/SchillingWNM18}
\bibfield{author}{\bibinfo{person}{Robert Schilling}, \bibinfo{person}{Mario
  Werner}, \bibinfo{person}{Pascal Nasahl}, {and} \bibinfo{person}{Stefan
  Mangard}.} \bibinfo{year}{2018}\natexlab{}.
\newblock
  \showarticletitle{\href{https://doi.org/10.1145/3274694.3274728}{Pointing in
  the Right Direction - Securing Memory Accesses in a Faulty World}}. In
  \bibinfo{booktitle}{\emph{Annual Computer Security Applications Conference --
  {ACSAC}}}.
\newblock


\bibitem[\protect\citeauthoryear{Serebryany}{Serebryany}{2019}]%
        {DBLP:journals/usenix-login/Serebryany19}
\bibfield{author}{\bibinfo{person}{Kostya Serebryany}.}
  \bibinfo{year}{2019}\natexlab{}.
\newblock
  \showarticletitle{\href{https://www.usenix.org/publications/login/summer2019/serebryany}{{ARM}
  Memory Tagging Extension and How It Improves {C/C++} Memory Safety}}.
\newblock \bibinfo{journal}{\emph{login Usenix Mag.}}  \bibinfo{volume}{44}
  (\bibinfo{year}{2019}).
\newblock


\bibitem[\protect\citeauthoryear{Serebryany, Bruening, Potapenko, and
  Vyukov}{Serebryany et~al\mbox{.}}{2012}]%
        {DBLP:conf/usenix/SerebryanyBPV12}
\bibfield{author}{\bibinfo{person}{Konstantin Serebryany},
  \bibinfo{person}{Derek Bruening}, \bibinfo{person}{Alexander Potapenko},
  {and} \bibinfo{person}{Dmitriy Vyukov}.} \bibinfo{year}{2012}\natexlab{}.
\newblock
  \showarticletitle{\href{https://www.usenix.org/conference/atc12/technical-sessions/presentation/serebryany}{AddressSanitizer:
  {A} Fast Address Sanity Checker}}. In \bibinfo{booktitle}{\emph{{USENIX}
  Annual Technical Conference}}.
\newblock


\bibitem[\protect\citeauthoryear{Serebryany and {Herle, Sudhi}}{Serebryany and
  {Herle, Sudhi}}{2019}]%
        {Serebryany2019}
\bibfield{author}{\bibinfo{person}{Kostya Serebryany} {and}
  \bibinfo{person}{{Herle, Sudhi}}.} \bibinfo{year}{2019}\natexlab{}.
\newblock
  \bibinfo{booktitle}{\emph{\href{https://security.googleblog.com/2019/08/adopting-arm-memory-tagging-extension.html}{{Adopting
  the Arm Memory Tagging Extension in Android}}}}.
\newblock


\bibitem[\protect\citeauthoryear{Serebryany, Stepanov, Shlyapnikov,
  Tsyrklevich, and Vyukov}{Serebryany et~al\mbox{.}}{2018}]%
        {DBLP:journals/corr/abs-1802-09517}
\bibfield{author}{\bibinfo{person}{Kostya Serebryany}, \bibinfo{person}{Evgenii
  Stepanov}, \bibinfo{person}{Aleksey Shlyapnikov}, \bibinfo{person}{Vlad
  Tsyrklevich}, {and} \bibinfo{person}{Dmitry Vyukov}.}
  \bibinfo{year}{2018}\natexlab{}.
\newblock \showarticletitle{\href{http://arxiv.org/abs/1802.09517}{Memory
  Tagging and how it improves {C/C++} memory safety}}.
\newblock \bibinfo{journal}{\emph{arXiv abs/1802.09517}}
  (\bibinfo{year}{2018}).
\newblock


\bibitem[\protect\citeauthoryear{Shacham}{Shacham}{2007}]%
        {DBLP:conf/ccs/Shacham07}
\bibfield{author}{\bibinfo{person}{Hovav Shacham}.}
  \bibinfo{year}{2007}\natexlab{}.
\newblock \showarticletitle{\href{https://doi.org/10.1145/1315245.1315313}{The
  geometry of innocent flesh on the bone: return-into-libc without function
  calls (on the x86)}}. In \bibinfo{booktitle}{\emph{Conference on Computer and
  Communications Security -- {CCS}}}.
\newblock


\bibitem[\protect\citeauthoryear{Song, Moon, Alam, Yun, Lee, Kim, Lee, and
  Paek}{Song et~al\mbox{.}}{2016}]%
        {DBLP:conf/sp/SongMAYLKLP16}
\bibfield{author}{\bibinfo{person}{Chengyu Song}, \bibinfo{person}{Hyungon
  Moon}, \bibinfo{person}{Monjur Alam}, \bibinfo{person}{Insu Yun},
  \bibinfo{person}{Byoungyoung Lee}, \bibinfo{person}{Taesoo Kim},
  \bibinfo{person}{Wenke Lee}, {and} \bibinfo{person}{Yunheung Paek}.}
  \bibinfo{year}{2016}\natexlab{}.
\newblock \showarticletitle{\href{https://doi.org/10.1109/SP.2016.9}{{HDFI:}
  Hardware-Assisted Data-Flow Isolation}}. In \bibinfo{booktitle}{\emph{{IEEE}
  Symposium on Security and Privacy -- {S{\&}P}}}.
\newblock


\bibitem[\protect\citeauthoryear{Song, Bradbury, and Mullins}{Song
  et~al\mbox{.}}{2015}]%
        {song2015towards}
\bibfield{author}{\bibinfo{person}{Wei Song}, \bibinfo{person}{Alex Bradbury},
  {and} \bibinfo{person}{Robert Mullins}.} \bibinfo{year}{2015}\natexlab{}.
\newblock \showarticletitle{Towards general purpose tagged memory}. In
  \bibinfo{booktitle}{\emph{Proceedings of the RISC-V Workshop}}.
\newblock


\bibitem[\protect\citeauthoryear{Szekeres, Payer, Wei, and Song}{Szekeres
  et~al\mbox{.}}{2013}]%
        {DBLP:conf/sp/SzekeresPWS13}
\bibfield{author}{\bibinfo{person}{Laszlo Szekeres}, \bibinfo{person}{Mathias
  Payer}, \bibinfo{person}{Tao Wei}, {and} \bibinfo{person}{Dawn Song}.}
  \bibinfo{year}{2013}\natexlab{}.
\newblock \showarticletitle{\href{https://doi.org/10.1109/SP.2013.13}{SoK:
  Eternal War in Memory}}. In \bibinfo{booktitle}{\emph{{IEEE} Symposium on
  Security and Privacy -- {S{\&}P}}}.
\newblock


\bibitem[\protect\citeauthoryear{Team}{Team}{2020a}]%
        {ASAN2020}
\bibfield{author}{\bibinfo{person}{The~Clang Team}.}
  \bibinfo{year}{2020}\natexlab{a}.
\newblock
  \bibinfo{booktitle}{\emph{\href{https://clang.llvm.org/docs/AddressSanitizer.html}{AddressSanitizer}}}.
\newblock


\bibitem[\protect\citeauthoryear{Team}{Team}{2020b}]%
        {HWASAN2020}
\bibfield{author}{\bibinfo{person}{The~Clang Team}.}
  \bibinfo{year}{2020}\natexlab{b}.
\newblock
  \bibinfo{booktitle}{\emph{\href{https://clang.llvm.org/docs/HardwareAssistedAddressSanitizerDesign.html}{Hardware-assisted
  AddressSanitizer Design Documentation}}}.
\newblock


\bibitem[\protect\citeauthoryear{Unterluggauer, Werner, and
  Mangard}{Unterluggauer et~al\mbox{.}}{2019}]%
        {DBLP:journals/jce/UnterluggauerWM19}
\bibfield{author}{\bibinfo{person}{Thomas Unterluggauer},
  \bibinfo{person}{Mario Werner}, {and} \bibinfo{person}{Stefan Mangard}.}
  \bibinfo{year}{2019}\natexlab{}.
\newblock
  \showarticletitle{\href{https://doi.org/10.1007/s13389-018-0180-2}{{MEAS:}
  memory encryption and authentication secure against side-channel attacks}}.
\newblock \bibinfo{journal}{\emph{J. Cryptographic Engineering}}
  \bibinfo{volume}{9} (\bibinfo{year}{2019}).
\newblock


\bibitem[\protect\citeauthoryear{Venkataramani, Doudalis, Solihin, and
  Prvulovic}{Venkataramani et~al\mbox{.}}{2008}]%
        {DBLP:conf/hpca/VenkataramaniDSP08}
\bibfield{author}{\bibinfo{person}{Guru Venkataramani},
  \bibinfo{person}{Ioannis Doudalis}, \bibinfo{person}{Yan Solihin}, {and}
  \bibinfo{person}{Milos Prvulovic}.} \bibinfo{year}{2008}\natexlab{}.
\newblock
  \showarticletitle{\href{https://doi.org/10.1109/HPCA.2008.4658637}{FlexiTaint:
  {A} programmable accelerator for dynamic taint propagation}}. In
  \bibinfo{booktitle}{\emph{International Conference on High-Performance
  Computer Architecture -- {HPCA}}}.
\newblock


\bibitem[\protect\citeauthoryear{Waterman, Lee, Avizienis, Patterson, and
  Asanovi\'{c}}{Waterman et~al\mbox{.}}{2016}]%
        {NON:waterman_riscvpriv191_2016}
\bibfield{author}{\bibinfo{person}{Andrew Waterman}, \bibinfo{person}{Yunsup
  Lee}, \bibinfo{person}{Rimas Avizienis}, \bibinfo{person}{David~A.
  Patterson}, {and} \bibinfo{person}{Krste Asanovi\'{c}}.}
  \bibinfo{year}{2016}\natexlab{}.
\newblock
  \bibinfo{booktitle}{\emph{\href{http://www2.eecs.berkeley.edu/Pubs/TechRpts/2016/EECS-2016-161.pdf}{{The
  RISC-V Instruction Set Manual Volume II: Privileged Architecture Version
  1.9.1}}}}.
\newblock \bibinfo{type}{{T}echnical {R}eport}. \bibinfo{institution}{EECS
  Department, University of California, Berkeley}.
\newblock


\bibitem[\protect\citeauthoryear{Weiser, Werner, Brasser, Malenko, Mangard, and
  Sadeghi}{Weiser et~al\mbox{.}}{2019}]%
        {DBLP:conf/ndss/WeiserWBMMS19}
\bibfield{author}{\bibinfo{person}{Samuel Weiser}, \bibinfo{person}{Mario
  Werner}, \bibinfo{person}{Ferdinand Brasser}, \bibinfo{person}{Maja Malenko},
  \bibinfo{person}{Stefan Mangard}, {and} \bibinfo{person}{Ahmad{-}Reza
  Sadeghi}.} \bibinfo{year}{2019}\natexlab{}.
\newblock
  \showarticletitle{\href{https://www.ndss-symposium.org/ndss-paper/timber-v-tag-isolated-memory-bringing-fine-grained-enclaves-to-risc-v/}{{TIMBER-V:}
  Tag-Isolated Memory Bringing Fine-grained Enclaves to {RISC-V}}}. In
  \bibinfo{booktitle}{\emph{Network and Distributed System Security Symposium
  -- {NDSS}}}.
\newblock


\bibitem[\protect\citeauthoryear{Werner, Unterluggauer, Schilling,
  Schaffenrath, and Mangard}{Werner et~al\mbox{.}}{2017}]%
        {DBLP:conf/fpl/WernerUSSM17}
\bibfield{author}{\bibinfo{person}{Mario Werner}, \bibinfo{person}{Thomas
  Unterluggauer}, \bibinfo{person}{Robert Schilling}, \bibinfo{person}{David
  Schaffenrath}, {and} \bibinfo{person}{Stefan Mangard}.}
  \bibinfo{year}{2017}\natexlab{}.
\newblock
  \showarticletitle{\href{https://doi.org/10.23919/FPL.2017.8056797}{Transparent
  memory encryption and authentication}}. In \bibinfo{booktitle}{\emph{Field
  Programmable Logic and Applications -- {FPL}}}.
\newblock


\bibitem[\protect\citeauthoryear{Woodruff, Watson, Chisnall, Moore, Anderson,
  Davis, Laurie, Neumann, Norton, and Roe}{Woodruff et~al\mbox{.}}{2014}]%
        {DBLP:conf/isca/WoodruffWCMADLNNR14}
\bibfield{author}{\bibinfo{person}{Jonathan Woodruff}, \bibinfo{person}{Robert
  N.~M. Watson}, \bibinfo{person}{David Chisnall}, \bibinfo{person}{Simon~W.
  Moore}, \bibinfo{person}{Jonathan Anderson}, \bibinfo{person}{Brooks Davis},
  \bibinfo{person}{Ben Laurie}, \bibinfo{person}{Peter~G. Neumann},
  \bibinfo{person}{Robert~M. Norton}, {and} \bibinfo{person}{Michael Roe}.}
  \bibinfo{year}{2014}\natexlab{}.
\newblock
  \showarticletitle{\href{https://doi.org/10.1109/ISCA.2014.6853201}{The
  {CHERI} capability model: Revisiting {RISC} in an age of risk}}. In
  \bibinfo{booktitle}{\emph{International Symposium on Computer Architecture --
  {ISCA}}}.
\newblock


\bibitem[\protect\citeauthoryear{Xu, DuVarney, and Sekar}{Xu
  et~al\mbox{.}}{2004}]%
        {xu2004efficient}
\bibfield{author}{\bibinfo{person}{Wei Xu}, \bibinfo{person}{Daniel~C
  DuVarney}, {and} \bibinfo{person}{R Sekar}.} \bibinfo{year}{2004}\natexlab{}.
\newblock \showarticletitle{An efficient and backwards-compatible
  transformation to ensure memory safety of C programs}. In
  \bibinfo{booktitle}{\emph{Proceedings of the 12th ACM SIGSOFT twelfth
  international symposium on Foundations of software engineering}}.
\newblock


\bibitem[\protect\citeauthoryear{Zaruba and Benini}{Zaruba and Benini}{2019}]%
        {DBLP:journals/tvlsi/ZarubaB19}
\bibfield{author}{\bibinfo{person}{Florian Zaruba} {and} \bibinfo{person}{Luca
  Benini}.} \bibinfo{year}{2019}\natexlab{}.
\newblock
  \showarticletitle{\href{https://doi.org/10.1109/TVLSI.2019.2926114}{The Cost
  of Application-Class Processing: Energy and Performance Analysis of a
  Linux-Ready 1.7-GHz 64-Bit {RISC-V} Core in 22-nm {FDSOI} Technology}}.
\newblock \bibinfo{journal}{\emph{{IEEE} Trans. Very Large Scale Integr.
  Syst.}}  \bibinfo{volume}{27} (\bibinfo{year}{2019}).
\newblock


\bibitem[\protect\citeauthoryear{Zeldovich, Kannan, Dalton, and
  Kozyrakis}{Zeldovich et~al\mbox{.}}{2008}]%
        {DBLP:conf/osdi/ZeldovichKDK08}
\bibfield{author}{\bibinfo{person}{Nickolai Zeldovich}, \bibinfo{person}{Hari
  Kannan}, \bibinfo{person}{Michael Dalton}, {and} \bibinfo{person}{Christos
  Kozyrakis}.} \bibinfo{year}{2008}\natexlab{}.
\newblock
  \showarticletitle{\href{http://www.usenix.org/events/osdi08/tech/full_papers/zeldovich/zeldovich.pdf}{Hardware
  Enforcement of Application Security Policies Using Tagged Memory}}. In
  \bibinfo{booktitle}{\emph{{USENIX} Symposium on Operating Systems Design and
  Implementation -- {OSDI}}}.
\newblock


\end{thebibliography}

\end{document}